\magnification\magstep 1
\parskip 5pt plus 1pt minus 0.5pt
\font\tbfi=cmmib10
\font\tenbi=cmmib7
\font\fivebi=cmmib5
\textfont4=\tbfi
\scriptfont4=\tenbi
\scriptscriptfont4=\fivebi
\mathchardef\gammac="700D

\mathchardef\thetac="7112
\def\thetab{{\fam4\thetac}}
\mathchardef\varphic="7127
\def\varphib{{\fam4\varphic}}
\def\dbl#1#2{\matrix{\hbox{#1}\cr\hbox{#2}\cr}}
\def\inner#1#2#3{\bigl\langle{#1}\big\vert{#2}\big\vert{#3}\bigr\rangle}
\def\init{\tabskip 0pt}
\def\crr{\cr\noalign{\hrule}}
\def\frd#1#2{{\displaystyle{\displaystyle#1\over\displaystyle#2}}}
\def\frac#1#2{{#1\over#2}}
\def\G{{\bf\Gamma}}
\def\g{\gamma}
\def\GH{{{\bf\Gamma}_H}}
\def\Gr{{\bf G}}
\def\A{{\bf A}}
\def\E{{\bf E}}
\def\I{{\bf I}}
\def\nat{_{\rm nat}}
\def\sca{_{\rm scal}}
\def\dif{^{\rm diff}}
\def\K{{\bf K}}
\def\L{{\bf L}}
\def\M{{\bf M}}
\def\P{{\bf P}}
\def\Q{{\bf Q}}
\def\n{{\bf n}}
\def\q{{\bf q}}
\def\R{{\bf R}}
\def\T{{\bf T}}
\def\1{{\bf 1}}
\def\s{\sigma}
\def\d{{\rm d}}
\def\p{\partial}
\def\i#1{^{(#1)}}
\def\ir#1{^{(\rm #1)}}
\def\vec#1{\widehat{\bf #1}}
\def\w{\widetilde}
\def\re{\mathop{\rm Re}}
\def\im{\mathop{\rm Im}}
\def\ca#1{{\cal #1}}
\def\l{{\Vert}}
\def\r{{\perp}}
\centerline{\bf MULTIPLE RAYLEIGH SCATTERING}
\smallskip
\centerline{\bf OF ELECTROMAGNETIC WAVES}
\bigskip
\centerline{\bf(revised version)}
\bigskip\null\medskip
\centerline{by E. Amic$^{(1)}$, J.M. Luck$^{(2)}$,}
\smallskip
\centerline{C.E.A. Saclay, Service de Physique Th\'eorique,
91191 Gif-sur-Yvette cedex, France,}
\bigskip
\centerline{and Th.M. Nieuwenhuizen$^{(3)}$,}
\smallskip
\centerline{Van der Waals-Zeeman Laboratorium,}
\centerline{Valckenierstraat 65, 1018 XE Amsterdam, The Netherlands.}
\vfill
\noindent{\bf Abstract.}
Multiple scattering of polarised electromagnetic waves in diffusive media
is investigated by means of radiative transfer theory.
This approach amounts to summing the ladder diagrams
for the diffuse reflected or transmitted intensity,
or the cyclical ones for the cone of enhanced backscattering.
The method becomes exact in several situations of interest,
such as a thick-slab experiment
(slab thickness $L\gg$ mean free path $\ell\gg$ wavelength $\lambda)$.
The present study is restricted to Rayleigh scattering.
It incorporates in a natural way the dependence on
the incident and detected polarisations,
and takes full account of the internal reflections at the boundaries of the
sample, due to the possible mismatch between the mean optical index $n$
of the medium and that $n_1$ of the surroundings.
This work does not rely on the diffusion approximation.
It therefore correctly describes radiation in the skin layers,
where a crossover takes place between free and diffusive propagation, and
vice-versa.
Quantities of interest, such as the polarisation-dependent,
angle-resolved mean diffuse intensity in reflection and in transmission,
and the shape of the cone of enhanced backscattering,
are predicted in terms of solutions to Schwarzschild-Milne equations.
The latter are obtained analytically, both in the absence of internal
reflections $(n=n_1)$,
and in the regime of a large index mismatch $(n/n_1\ll 1$ or $\gg 1$).
\vfill
{\parskip 0pt
\noindent(1) e-mail: amic@spht.saclay.cea.fr

\noindent(2) e-mail: luck@spht.saclay.cea.fr

\noindent(3) e-mail: nieuwenh@phys.uva.nl
}
\eject
\noindent{\bf 1. INTRODUCTION}

Light undergoes multiple scattering when propagating through inhomogeneous
media over distances much larger than one mean free path $\ell$.
This may occur in a wide variety of situations,
ranging from the atmospheres of stars and planets to biological tissues.
The theory of multiple scattering of electromagnetic waves
is an old classical area of physics [1--4],
which has been developed for almost one century, mostly by astrophysicists.
This subject has been experiencing an important revival
of theoretical and experimental activity for one decade,
motivated by the analogy between the effects of random disorder
(weak or strong localisation)
on the propagation of classical waves (electromagnetic, acoustic, seismic)
and of quantum-mechanical waves (electrons in solids).
The first weak-localisation effect to be discovered
has been the celebrated enhanced backscattering phenomenon,
which takes place in a narrow angular cone around the direction of
exact backscattering [5].

Typical laboratory experiments on multiple light scattering
involve suspensions of polystyrene spheres or of TiO$_2$
(white paint) grains in fluids.
In these situations the mean free path $\ell$ is usually
much larger than the wavelength $\lambda$ of light,
and the samples are often optically thick slabs, of thickness $L\gg\ell$.
The regime of interest, i.e., $\lambda\ll\ell\ll L$,
is characterised by a diffusive transport of radiation
through multiple scattering.
This diffusive regime admits three different levels of theoretical description.
(i) The crudest approach is the diffusion approximation,
where the multiply scattered intensity is described
by means of an effective diffusion equation.
The latter is only valid on length scales larger than $\ell$,
so that is has to be supplemented by boundary conditions.
As a consequence, this approach somehow keeps a phenomenological character.
(ii) The mesoscopic approach, known as radiative transfer theory (RTT),
has been used for long by the community of astrophysicists [1--4].
It is based on a local balance equation, keeping track of the direction
of propagation of the intensity.
(iii) The systematic microscopic approach consists in expanding
the solution of the Maxwell equations in the random medium
as a diagrammatic Born (multiple-scattering) series.

In multiple light scattering experiments the quantities of most interest
are the mean diffuse reflected and transmitted intensity,
and the shape of the peak of enhanced backscattering.
For these observables the diagrammatic approach greatly simplifies
in the regime $\lambda\ll\ell\ll L$.
As far as mean quantities, averaged over the random positions
of the scatterers, are concerned,
the diffuse radiation is described by the sum of the ladder diagrams,
which, in turn, amounts to RTT;
the enhanced backscattering phenomenon is
described by the so-called cyclical or maximally-crossed diagrams,
which can also be summed up by an adaptation of RTT [6, 7].
On the other hand, the validity of vector RTT
has been established on a rigorous basis [8],
starting from a perturbative treatment of Maxwell's equations,
extending earlier developments [9] on multiple scattering of
electromagnetic waves in plasmas.
In the regime where the random fluctuations of the dielectric constant
have short-range correlations, this approach rigorously justifies the use of
the Schwarzschild-Milne equation of vector RTT
with the Rayleigh phase function,
which will be the purpose of the present work.

The principles of vector RTT
for electromagnetic waves, taking into account their polarisations,
are exposed in the book by Chandrasekhar [1],
which also contains a formal analytical derivation of the diffuse intensity
for Rayleigh scattering, in the absence of internal reflections.
This approach is needed in order to obtain predictions
at a quantitative level, concerning observables like the diffuse intensity
in reflection and in transmission, and the enhanced backscattering cone.
In particular the diffusion approximation alone cannot yield such accurate
predictions,
chiefly because boundary conditions cannot be dealt with
in a fully satisfactory way.
Surprisingly enough, in the modern era of multiple scattering
only very few authors have used RTT.
Thus far the major investigations of the weak localisation
of light, including polarisation effects,
have rather used either the diffusion approximation [10--12]
or numerical simulations [13].
Several other bulk properties of multiple scattering of electromagnetic waves
have been investigated along these lines, including especially
the effects of Faraday rotation [12, 14, 15] and of absorption [16].
Exact results on polarisation effects on the backscattering cone
have only appeared very recently.
Mishchenko [17] has derived general properties
of the behaviour of polarisations under time reversal,
obtaining thus for the first time a consistent derivation of
the enhancement factors in the direction of exact backscattering.
The full shape of the backscattering cone has then been investigated by Ozrin
[18], who did not, however, come up with a full analytical solution
of the latter problem.
In previous works, we have considered the case of scalar waves
undergoing multiple isotropic [19, 20] and arbitrary anisotropic scattering
[21].
We have shown how RTT takes proper account of the skin layers,
where light is converted over a few mean free paths from a free beam
to a diffusive field and vice-versa,
and how it allows to deal with the effects of internal reflections
due to the index mismatch at the boundaries of the sample.
The latter effect has been the subject of much activity recently [22--26].

The goal of the present paper is to extend our investigations
to the multiple scattering of electromagnetic waves,
obtaining thus for the first time a complete analytic description
of the diffuse intensity and of the backscattering cone
in the regime $\lambda\ll\ell\ll L$,
including both polarisation effects and internal reflections.
We shall restrict the analysis to Rayleigh scatterers for definiteness.
Section 2 contains general results on vector RTT.
The observables of interest, with their dependence on polarisations
and index mismatch, are expressed in terms of solutions of appropriate
Schwarzschild-Milne equations.
Sum rules and other general properties are given.
These predictions are then made more quantitative in two situations:
(a) in the absence of internal reflections (section 3),
where a full analytical solution of the problem is given;
the results of refs. [1, 17, 18] are made more precise
and put in a broader perspective;
(b) in the opposite regime of a large mismatch of optical index
between the sample and the surroundings (section 4).
Section 5 contains a brief discussion of our findings.
\smallskip
\noindent{\bf 2. GENERALITIES}

\noindent{\bf 2.1. General formalism}

Throughout the following we consider
the multiple scattering of electromagnetic waves
by a diffusive medium containing a low density $\rho$ of identical
Rayleigh scatterers characterised by their cross-section $\s$,
so that the mean free path $\ell=1/(\rho\s)$ is much larger
than the wavelength $\lambda$ of radiation in the medium.
The diffusive medium has the form of a slab $(0<z<L)$,
infinite in the transverse directions.
We introduce the optical thickness $b$ of the sample through $L=b\ell$,
and the optical depth $\tau$ of a point in the sample
$(0<\tau<b)$ through $z=\tau\ell$.
We shall consider either optically thick slabs $(b\gg 1)$,
or semi-infinite samples $(b=+\infty)$.
We investigate the general situation where the mean optical index $n$
of the sample is different from that $n_1$ of the surrounding medium.
Whenever there is an index mismatch $(m=n/n_1\ne 1)$,
internal reflections take place at the boundaries of the sample.
Useful definitions and notations are summarised in Table 1.

We closely follow the definitions and notations of Chandrasekhar [1].
We measure the polarisation of the radiation in a fixed frame,
introducing spherical co-ordinates (see Table 1).
For radiation propagating in the angular direction $(\theta,\varphi)$
with respect to the $z$-axis,
the complex components of the electric field $\E$
in the plane transversal to that direction will be denoted by $(E_\theta,
E_\varphi)$.
The component $E_\theta$ is parallel to the unit vector $\vec{\thetab}$
and contained in the meridian plane,
defined by the direction of propagation and the normal to the boundaries
of the sample, while the component $E_\varphi$ is parallel to the unit
vector $\vec{\varphib}$ and normal to the meridian plane.
The alternative notations $(E_\l, E_\r)$ and $(E_\ell, E_r)$
can be found in the literature.
We introduce the following vector of four Stokes parameters, or Stokes vector
for short
$$
\I:\left\vert\matrix{
\I_1=\vert E_\theta\vert^2\hfill\cr
\I_2=\vert E_\varphi\vert^2\hfill\cr
\I_3=U=2\re\big(E_\theta E_\varphi^*\big)\hfill\cr
\I_4=V=2\im\big(E_\theta E_\varphi^*\big).\hfill\cr
}\right.
\eqno(2.1)
$$
Here and throughout the following,
boldface symbols represent 4-component vectors or $4\times 4$ matrices.
The description of polarised radiation by means of Stokes parameters
is very commonly used [27].
This formalism has many advantages:
the Stokes parameters add up for light beams superposed incoherently;
a scattering event is described by a linear transformation
of the Stokes parameters, i.e., by the action of
a scattering matrix on the Stokes vector $\I$.
We finally recall the definition [27] of the degree of polarisation $P$
of radiation described by a Stokes vector $\I$:
$$
P={\sqrt{(\I_1-\I_2)^2+U^2+V^2}\over\I_1+\I_2}
.\eqno(2.2)
$$
\smallskip
\noindent{\bf 2.2. Schwarzschild-Milne equation}

For reasons exposed in the Introduction, we shall use RTT
to investigate the average reflected or transmitted intensity
in the regime $\ell\gg\lambda$.
We consider first the situation of a semi-infinite medium, for simplicity.

The mean diffuse radiation propagating in the direction $(\theta,\varphi)$
in the medium at depth $\tau=z/\ell$ is described by its
Stokes vector $\I(\tau,\mu,\varphi)$, with the notations (see Table 1)
$$
\mu=\cos\theta,\quad\nu=\sin\theta=\sqrt{1-\mu^2}
.\eqno(2.3)
$$
Along the lines of ref. [1], the RTT equation reads
$$
\mu{\p\over\p\tau}\I(\tau,\mu,\varphi)
=\G(\tau,\mu,\varphi)-\I(\tau,\mu,\varphi)
,\eqno(2.4)
$$
where the vector source function $\G(\tau,\mu,\varphi)$ is defined as
$$
\G(\tau,\mu,\varphi)
=\int_{-1}^1{\d\mu'\over2\mu'}\int_0^{2\pi}{\d\varphi'\over2\pi}
\P(\mu,\varphi,\mu',\varphi').\I(\tau,\mu',\varphi')
,\eqno(2.5)
$$
with the Rayleigh phase matrix $\P(\mu,\varphi,\mu',\varphi')$
being the matrix describing a scattering event,
expressed in the fixed frame related to the sample.
Its explicit form will be given in eqs. (2.10), (2.11).

The RTT equation (2.4), with appropriate boundary conditions,
leads to the following linear integral equation for the source function,
referred to as the Schwarzschild-Milne (SM) equation
$$
\eqalign{
\G(\tau,\mu,\varphi)
&=\P(\mu,\varphi,\mu_a,\varphi_a).\I_ae^{-\tau/\mu_a}\cr
&+\int_0^\tau\d\tau'\int_0^1{\d\mu'\over2\mu'}\int_0^{2\pi}{\d\varphi'\over
2\pi}
e^{-(\tau-\tau')/\mu'}\P(\mu,\varphi,\mu',\varphi').\G(\tau',\mu',\varphi')\cr
&+\int_\tau^{+\infty}\d\tau'\int_0^1{\d\mu'\over2\mu'}
\int_0^{2\pi}{\d\varphi'\over 2\pi}e^{-(\tau'-\tau)/\mu'}
\P(\mu,\varphi,-\mu',\varphi').\G(\tau',-\mu',\varphi')\cr
&+\int_0^{+\infty}\d\tau'\int_0^1{\d\mu'\over2\mu'}
\int_0^{2\pi}{\d\varphi'\over 2\pi}
e^{-(\tau+\tau')/\mu'}\R(\mu').\P(\mu,\varphi,\mu',\varphi').
\G(\tau',\mu',\varphi').\cr
}
\eqno(2.6)
$$
The right-hand side of this equation has the following interpretation:

\noindent$\bullet$
The first line is the exponentially damped contribution,
with a suitable normalisation [19, 20],
of the incident beam,
characterised by an incident direction $(\theta_a,\varphi_a)$
and a Stokes vector $\I_a$;

\noindent$\bullet$
The second (third) line is the {\it bulk} contribution
of diffuse light scattered from a smaller (larger) depth $\tau'$;

\noindent$\bullet$
The fourth line is the {\it layer} contribution of diffuse light
scattered from depth $\tau'$ and then reflected at the boundary $(\tau=0)$.
The effect of the boundary is described by a reflection matrix $\R(\mu)$
and a transmission matrix $\T(\mu)$, namely
$$
\eqalign{
\R(\mu)&=\pmatrix{
\big\vert r_\l(\mu)\big\vert^2 &0&0&0\cr
0&\big\vert r_\r(\mu)\big\vert^2 &0&0\cr
0&0&\re\bigl(r_\l(\mu)r_\r(\mu)^*\bigr)&-\im\bigl(r_\l(\mu)r_\r(\mu)^*\bigr)\cr
0&0&\im\bigl(r_\l(\mu)r_\r(\mu)^*\bigr)&\re\bigl(r_\l(\mu)r_\r(\mu)^*\bigr)\cr
},\cr\cr
\T(\mu)&={m\mu\over\sqrt{1-m^2\nu^2}}\pmatrix{
\big\vert t_\l(\mu)\big\vert^2 &0&0&0\cr
0&\big\vert t_\r(\mu)\big\vert^2 &0&0\cr
0&0&\re\bigl(t_\l(\mu)t_\r(\mu)^*\bigr)&-\im\bigl(t_\l(\mu)t_\r(\mu)^*\bigr)\cr
0&0&\im\bigl(t_\l(\mu)t_\r(\mu)^*\bigr)&\re\bigl(t_\l(\mu)t_\r(\mu)^*\bigr)\cr
}.\cr
}
\eqno(2.7)
$$
In these expressions $r_\l(\mu)$, $r_\r(\mu)$ and $t_\l(\mu)$, $t_\r(\mu)$
are the Fresnel reflection and transmission amplitude coefficients,
respectively.
The latter only depend on the inner incidence angle $\theta$
and on the index mismatch $m$, according to
$$
\eqalign{
r_\l(\mu)&={\mu-m\sqrt{1-m^2\nu^2}\over\mu+m\sqrt{1-m^2\nu^2}},\quad
r_\r(\mu)={\sqrt{1-m^2\nu^2}-m\mu\over\sqrt{1-m^2\nu^2}+m\mu},\cr
t_\l(\mu)&={2\sqrt{1-m^2\nu^2}\over\mu+m\sqrt{1-m^2\nu^2}},\quad
t_\r(\mu)={2\sqrt{1-m^2\nu^2}\over\sqrt{1-m^2\nu^2}+m\mu}.\cr
}
\eqno(2.8)
$$
In the case of partial reflection (see Table 1),
these coefficients are real, with absolute values less than unity.
In the case of total reflection,
the reflection coefficients are pure phases, i.e., complex numbers with unit
modulus, while the transmission coefficients vanish by convention.
The first two diagonal elements of the reflection matrix $\R(\mu)$ of
eq. (2.7) read
$$
\big\vert r_\l(\mu)\big\vert^2=R_\l(\mu)=1-T_\l(\mu),\quad
\big\vert r_\r(\mu)\big\vert^2=R_\r(\mu)=1-T_\r(\mu)
,\eqno(2.9)
$$
in terms of the Fresnel reflection and transmission intensity coefficients.

It is advantageous to expand the $\varphi$-dependence of all quantities in the
complex
trigonometric polynomials $\bigl\{e^{ik\varphi}\bigr\}$, with $-2\le k\le 2$.
We thus set
$$
\eqalign{
&\I(\mu,\varphi)=\sum_{k=-2}^2\I\i{k}(\mu)e^{ik\varphi},\quad
\G(\mu,\varphi)=\sum_{k=-2}^2\G\i{k}(\mu)e^{ik\varphi},\cr
&\P(\mu,\varphi,\mu',\varphi')
=\sum_{k=-2}^2\P\i{k}(\mu,\mu')e^{ik(\varphi-\varphi')}.
}
\eqno(2.10)
$$
The Rayleigh phase matrices read
$$
\eqalign{
&\P\i{0}(\mu,\mu')=\frac{3}{4}\pmatrix{
2(1-\mu^2)(1-\mu'^2)+\mu^2\mu'^2 &\mu^2 & 0 & 0\cr
\mu'^2 & 1 & 0 & 0\cr
0 & 0 & 0 & 0\cr
0 & 0 & 0 & 2\mu\mu'\cr},\null\cr
&\P\i{1}(\mu,\mu')={\P\i{-1}}^*(\mu,\mu')=\frac{3}{4}\nu\nu'\pmatrix{
2\mu\mu' & 0 & i\mu & 0\cr
0 & 0 & 0 & 0\cr
-2i\mu' & 0 & 1 & 0\cr
0 & 0 & 0 & 1\cr},\null\cr
&\P\i{2}(\mu,\mu')={\P\i{-2}}^*(\mu,\mu')=\frac{3}{8}\pmatrix{
\mu^2\mu'^2 & -\mu^2 & i\mu^2\mu' & 0\cr
-\mu'^2 & 1 & -i\mu' & 0\cr
-2i\mu\mu'^2 & 2i\mu & 2\mu\mu' & 0\cr
0 & 0 & 0 & 0\cr},
}
\eqno(2.11)
$$
and the SM equation (2.6) splits into the following five
decoupled integral equations $(-2\le k\le 2)$
$$
\eqalign{
\G\i{k}(\tau,\mu)
&=\P\i{k}(\mu,\mu_a).\I_ae^{-ik\varphi_a-\tau/\mu_a}\cr
&+\int_0^\tau\d\tau'\int_0^1{\d\mu'\over2\mu'}e^{-(\tau-\tau')/\mu'}
\P\i{k}(\mu,\mu').\G\i{k}(\tau',\mu')\cr
&+\int_\tau^{+\infty}\d\tau'\int_0^1{\d\mu'\over2\mu'}e^{-(\tau'-\tau)/\mu'}
\P\i{k}(\mu,-\mu').\G\i{k}(\tau',-\mu')\cr
&+\int_0^{+\infty}\d\tau'\int_0^1{\d\mu'\over2\mu'}e^{-(\tau+\tau')/\mu'}
\R(\mu').\P\i{k}(\mu,\mu').\G\i{k}(\tau',\mu').\cr
}
\eqno(2.12)
$$
\smallskip
\noindent{\bf 2.3. Solutions to the SM equation and sum rules}

We now turn to the analysis of the solutions to the SM equation (2.6).
We shall investigate their general symmetry properties,
and show that they obey some remarkable sum rules.

We start by introducing the matrix Green's function
$\Gr(\tau,\mu,\varphi,\tau',\mu',\varphi')$, defined as being the solution,
which remains bounded as either $\tau$ or $\tau'$ goes to infinity,
of the SM equation with a matrix $\delta$-function source term, namely
$$
\eqalign{
&\Gr(\tau,\mu,\varphi,\tau',\mu',\varphi')
=\P(\mu,\varphi,\mu',\varphi')\delta(\tau-\tau')\cr
&+\int_0^\tau\d\tau''\int_0^1{\d\mu''\over2\mu''}\int_0^{2\pi}
{\d\varphi''\over2\pi}
e^{-(\tau-\tau'')/\mu''}\P(\mu,\varphi,\mu'',\varphi'').
\Gr(\tau'',\mu'',\varphi'',\tau',\mu',\varphi')\cr
&+\int_\tau^{+\infty}\d\tau''\int_0^1{\d\mu''\over2\mu''}
\int_0^{2\pi}{\d\varphi''\over 2\pi}
e^{-(\tau''-\tau)/\mu''}\P(\mu,\varphi,-\mu'',\varphi'').
\Gr(\tau'',-\mu'',\varphi'',\tau',\mu',\varphi')\cr
&+\int_0^{+\infty}\d\tau''\int_0^1{\d\mu''\over2\mu''}\int_0^{2\pi}
{\d\varphi''\over2\pi}
e^{-(\tau+\tau'')/\mu''}\R(\mu'').\P(\mu,\varphi,\mu'',\varphi'').
\Gr(\tau'',\mu'',\varphi'',\tau',\mu',\varphi').\cr
}
\eqno(2.13)
$$
The source function $\G(\tau,\mu,\varphi)$, solution of the SM equation (2.6),
is then given by
$$
\G(\tau,\mu,\varphi)=\G(\tau,\mu,\varphi,\mu_a,\varphi_a).\I_a
,\eqno(2.14)
$$
where
$$
\G(\tau,\mu,\varphi,\mu_a,\varphi_a)
=\int_0^{+\infty}\d\tau'e^{-\tau'/\mu_a}
\Gr(\tau,\mu,\varphi,\tau',\mu_a,\varphi_a)
.\eqno(2.15)
$$
We also define the matrix of bistatic coefficients, or bistatic
matrix for short
$$
\eqalign{
\g(\mu_a,\varphi_a,\mu_b,\varphi_b)
&=\int_0^{+\infty}\d\tau
e^{-\tau/\mu_b}\G(\tau,-\mu_b,\varphi_b,\mu_a,\varphi_a)\cr
&=\int_0^{+\infty}\d\tau e^{-\tau/\mu_b}\int_0^{+\infty}\d\tau'e^{-\tau'/\mu_a}
\Gr(\tau,-\mu_b,\varphi_b,\tau',\mu_a,\varphi_a).
}
\eqno(2.16)
$$

The invariance of the Rayleigh scattering mechanism under time reversal
implies the following symmetry properties of the quantities defined so far.
We introduce the constant matrices
$$
\K=\pmatrix{1&0&0&0\cr0&1&0&0\cr0&0&{1\over2}&0\cr0&0&0&{1\over2}\cr},
\quad\L=\pmatrix{1&0&0&0\cr0&1&0&0\cr0&0&-{1\over2}&0\cr0&0&0&{1\over2}\cr}
,\eqno(2.17)
$$
the matrix $\K$ being denoted by $\Q^{-1}$ in ref. [1].
The Rayleigh phase matrix $\P(\mu,\varphi,\mu',\varphi')$ has the symmetry
property $(i,j=1,\cdots,4)$
$$
\eqalign{
(\K.\P)_{ij}(\mu,\varphi,\mu',\varphi')
&=(\K.\P)_{ji}(\mu',\varphi',\mu,\varphi),\cr
(\L.\P)_{ij}(\mu,\varphi,\mu',\varphi')
&=(\L.\P)_{ji}(-\mu',\varphi',-\mu,\varphi).
}
\eqno(2.18)
$$
It then follows from their definitions (2.13), (2.16)
that the matrix Green's function and the bistatic matrix obey the symmetry
relations $(i,j=1,\cdots,4)$
$$
\eqalign{
(\K.\Gr)_{ij}(\tau,\mu,\varphi,\tau',\mu',\varphi')
&=(\K.\Gr)_{ji}(\tau',\mu',\varphi',\tau,\mu,\varphi),\cr
(\L.\Gr)_{ij}(\tau,\mu,\varphi,\tau',\mu',\varphi')
&=(\L.\Gr)_{ji}(\tau',-\mu',\varphi',\tau,-\mu,\varphi),\cr
(\L.\g)_{ij}(\mu,\varphi,\mu',\varphi')
&=(\L.\g)_{ji}(\mu',\varphi',\mu,\varphi).
}
\eqno(2.19)
$$

We now investigate the asymptotic behaviour of quantities
deep inside the medium, i.e., for $\tau\to+\infty$.
It is expected on physical grounds that the diffusive medium
depolarises the incident radiation, so that both $\I(\tau,\mu,\varphi)$
and $\G(\tau,\mu,\varphi)$ become proportional to the Stokes vector
of natural (unpolarised) light, namely [27]
$$
\I\nat=\pmatrix{1\cr1\cr0\cr0}
.\eqno(2.20)
$$
This assertion will be made quantitative in section 3,
where the extinction lengths of the all other modes will be determined.
The asymptotic behaviour of the matrix solution
$\G(\tau,\mu,\varphi,\mu_a,\varphi_a)$ then assumes the form
$$
\G(\tau,\mu,\varphi,\mu_a,\varphi_a)\to\pmatrix{
\tau_1(\mu_a) &\tau_2(\mu_a) & 0 & 0\cr
\tau_1(\mu_a) &\tau_2(\mu_a) & 0 & 0\cr
0 & 0 & 0 & 0\cr
0 & 0 & 0 & 0\cr
}\quad(\tau\to+\infty)
.\eqno(2.21)
$$
Furthermore it should be noticed that the homogeneous SM equation
(2.6), (2.12),
without a source term, has a vector solution $\GH(\tau,\mu)$
in the $\varphi$-independent $(k=0)$ sector, growing linearly as
$\tau\to+\infty$.
We shall refer to the latter solution, normalised as
$$
\GH(\tau,\mu)\approx(\tau+\tau_0)\I\nat
,\eqno(2.22)
$$
as the homogeneous solution, for short.

The constant $\tau_0$ and the functions $\tau_1(\mu)$ and $\tau_2(\mu)$,
which show up in eqs. (2.22) and (2.21), respectively, are unknown so far.
It will be shown that they bear the full non-trivial dependence
of the physical observables on the index mismatch.
They will also be determined analytically in the absence
of internal reflections
(section 3) and in the large index mismatch regime (section 4).

For the time being, we pursue our investigation of general properties.
The special and homogeneous solutions to the SM equation
in the $k=0$ sector can be related among themselves as follows.
The column vectors of the matrix Green's function obeying eq. (2.13)
become proportional to the homogeneous solution,
as either $\tau$ or $\tau'$ goes to infinity:
$$
\lim_{\tau'\to+\infty}\Gr\i{0}_{ij}(\tau,\mu,\tau',\mu')
={1\over 2D}(\GH)_i(\tau,\mu)\quad(i,j=1,\cdots,4)
,\eqno(2.23)
$$
where the constant $D$ will be determined and interpreted in a while.
As a consequence of the above definitions, we have
$$
\tau_i(\mu_a)=\lim_{\mu_b\to+\infty}{\g\i{0}_{ij}(\mu_a,\mu_b)\over\mu_b}
={1\over 2D}\int_0^{+\infty}\d\tau e^{-\tau/\mu_a}(\GH)_i(\tau,-\mu_a)
\quad(i,j=1,2)
.\eqno(2.24)
$$

We end up by deriving two groups of sum rules
obeyed by the quantities defined above,
which are related to the $F$- and $K$-integrals,
with the notations of ref. [1].

\noindent$\bullet$
First, we consider the $F$-integral, defined as
$$
F(\tau)=\int_{-1}^1{\d\mu\over 2}\mu\int_0^{2\pi}{\d\varphi\over 2\pi}
\Big(\I_1(\tau,\mu,\varphi)+\I_2(\tau,\mu,\varphi)\Big)
.\eqno(2.25)
$$
The RTT equation (2.4) implies $\d F/\d\tau=0$,
expressing thus the conservation of the flux in the $z$-direction.
We consider first the $F$-integrals $F_1$ and $F_2$
associated with the special solutions $\G_{i1}$ and $\G_{i2}$, i.e.,
the first two column vectors of the matrix (2.15).
The $\tau\to+\infty$ limit determines $F_1=F_2=0$;
the $\tau=0$ values then yield the sum rules
$$
\eqalign{
&\int_0^1{\d\mu\over 2}\Bigl[T_\l(\mu)\g\i{0}_{11}(\mu,\mu_a)
+T_\r(\mu)\g\i{0}_{12}(\mu,\mu_a)\Bigr]=\mu_a,\cr
&\int_0^1{\d\mu\over 2}\Bigl[T_\l(\mu)\g\i{0}_{21}(\mu,\mu_a)
+T_\r(\mu)\g\i{0}_{22}(\mu,\mu_a)\Bigr]=\mu_a.\cr
}
\eqno(2.26)
$$
The $\mu_a\to+\infty$ limit of these equations, using eq. (2.24), yields
$$
\int_0^1{\d\mu\over 2}\Bigl[T_\l(\mu)\tau_1(\mu)+T_\r(\mu)\tau_2(\mu)\Bigr]=1
.\eqno(2.27)
$$
Similarly, we consider the $F$-integral $F_H$
associated with the homogeneous solution $\GH$.
This does not yield any independent sum rule,
but rather leads to the determination of the unknown constant $D$, namely
$$
D=\frac{1}{3}
,\eqno(2.28)
$$
to be interpreted as the dimensionless diffusion constant, i.e.,
$D_{\rm phys}=c\ell/3$ in physical units.

\noindent$\bullet$
Second, we consider the $K$-integral, defined as
$$
K(\tau)=\int_{-1}^1{\d\mu\over 2}\mu^2
\int_0^{2\pi}{\d\varphi\over
2\pi}\Big(\I_1(\tau,\mu,\varphi)+\I_2(\tau,\mu,\varphi)\Big)
.\eqno(2.29)
$$
The RTT equation (2.4) implies $\d K/\d\tau=-F$,
whence $K(\tau)=-F\tau+K_0$, with $K_0$ being a constant.
Along the lines of the above derivation, we obtain the sum rules
$$
\eqalign{
&\int_0^1{\d\mu\over 2}\mu
\Bigl[\big(1+R_\l(\mu)\big)\g\i{0}_{11}(\mu,\mu_a)
+\big(1+R_\r(\mu)\big)\g\i{0}_{12}(\mu,\mu_a)\Bigr]
=\frac{2}{3}\tau_1(\mu_a)+\mu_a^2,\cr
&\int_0^1{\d\mu\over 2}\mu
\Bigl[\big(1+R_\l(\mu)\big)\g\i{0}_{21}(\mu,\mu_a)
+\big(1+R_\r(\mu)\big)\g\i{0}_{22}(\mu,\mu_a)\Bigr]
=\frac{2}{3}\tau_2(\mu_a)+\mu_a^2,\cr
}
\eqno(2.30)
$$
and
$$
\int_0^1{\d\mu\over 2}\mu
\Bigl[\big(1+R_\l(\mu)\big)\tau_1(\mu)
+\big(1+R_\r(\mu)\big)\tau_2(\mu)\Bigr]=\tau_0
.\eqno(2.31)
$$
The sum rules (2.30), (2.31) express
that the multiple-scattering problem in a semi-infinite sample
is invariant if a finite slab of any thickness is added, or removed,
from the sample [1].
\smallskip
\noindent{\bf 2.4. Diffuse reflected intensity}

The diffuse reflected intensity for a semi-infinite sample can now
be calculated, along the lines of refs. [19--21].
The incident radiation is characterised by the direction $(\theta_a,\varphi_a)$
and the Stokes vector $\I_a$;
the reflected radiation is detected in the direction $(\theta_b,\varphi_b)$
and in a polarisation state characterised by the Stokes vector $\I_b$.
Our prediction for the mean reflected intensity per solid-angle element reads
$$
{\d R(a\to b)\over\d\Omega_b}
=A^R(\theta_a,\varphi_a,\theta_b,\varphi_b)
={\cos\theta_a\over 4\pi m^2\mu_a\mu_b}
\inner{\I_b}{\T(\mu_b).\L.\g(\mu_a,\varphi_a,\mu_b,\varphi_b).\T(\mu_a)}{\I_a}
.\eqno(2.32)
$$
In the absence of index mismatch we obtain the simpler expression
$$
A^R(\theta_a,\varphi_a,\theta_b,\varphi_b)
={1\over 4\pi\mu_b}\inner{\I_b}{\L.\g(\mu_a,\varphi_a,\mu_b,\varphi_b)}{\I_a}
.\eqno(2.33)
$$
\smallskip
\noindent{\bf 2.5. Diffuse transmitted intensity}

The diffuse transmitted intensity through an optically thick slab $(b\gg 1)$
can also be calculated along the lines of refs. [19--21].
A first step consists in building up the solution
$\G(b,\tau,\mu,\varphi,\mu_a,\varphi_a)$ of the SM equation
pertaining to the thick-slab geometry.
This solution can be expressed in terms of the solutions
$\G(\tau,\mu,\varphi,\mu_a,\varphi_a)$ and $\GH(\tau,\mu)$
pertaining to the semi-infinite geometry, by means of a matching procedure.
It turns out that only the $(1,2)$ sector of the matrix solution matters,
since all the other matrix elements are exponentially small
in the optical thickness $b$.
We thus get $(i,j=1,2)$
$$
\G_{ij}(b,\tau,\mu,\varphi,\mu_a,\varphi_a)\approx\left\{\matrix{
\G_{ij}(\tau,\mu,\varphi,\mu_a,\varphi_a)
-\displaystyle{\tau_i(\mu_a)\over b+2\tau_0}
(\GH)_i(\tau,\mu)\hfill&(\tau\hbox{ finite},\;b-\tau\gg 1),\hfill\cr\null\cr
\displaystyle{\tau_i(\mu_a)\over
b+2\tau_0}(\GH)_i(b-\tau,-\mu)\hfill&(b-\tau\hbox{ finite},\;\tau\gg 1).
\hfill\cr}\right.
\eqno(2.34)
$$
Both expressions lead to a linear (diffusive) behaviour
in the bulk of the sample ($\tau\gg 1$, $b-\tau\gg 1$), namely
$$
\G_{ij}(b,\tau,\mu,\varphi,\mu_a,\varphi_a)
\approx{b+\tau_0-\tau\over b+2\tau_0}\tau_i(\mu_a)\quad(i,j=1,2)
.\eqno(2.35)
$$

The mean diffuse transmitted intensity through an optically thick slab
can then be derived explicitly, again along the lines of refs. [19--21].
The incident radiation is characterised by the direction $(\theta_a,\varphi_a)$
and the Stokes vector $\I_a$.
The transmitted radiation is detected in the direction $(\theta_b,\varphi_b)$
and in a polarisation state characterised by the Stokes vector $\I_b$.
Our prediction for the mean transmitted intensity per solid-angle element reads
$$
{\d T(a\to b)\over\d\Omega_b}
={A^T(\theta_a,\theta_b)\over b+2\tau_0}
,\eqno(2.36)
$$
with
$$
A^T(\theta_a,\theta_b)={\cos\theta_a\over 12\pi m^2\mu_a\mu_b}
\inner{\I_b}{\A(\mu_a,\mu_b)}{\I_a}
.\eqno(2.37)
$$
The matrix $\A(\mu_a,\mu_b)$ has non-zero elements only
in the $(1,2)$-sector, namely
$$
\A(\mu_a,\mu_b)=\pmatrix{
T_\l(\mu_a)\tau_1(\mu_a)T_\l(\mu_b)\tau_1(\mu_b)&
T_\l(\mu_a)\tau_1(\mu_a)T_\r(\mu_b)\tau_2(\mu_b)&0&0\cr
T_\r(\mu_a)\tau_2(\mu_a)T_\l(\mu_b)\tau_1(\mu_b)&
T_\r(\mu_a)\tau_2(\mu_a)T_\r(\mu_b)\tau_2(\mu_b)&0&0\cr
0&0&0&0\cr 0&0&0&0\cr}
.\eqno(2.38)
$$
\smallskip
\noindent{\bf 2.6. Enhanced backscattering cone}

\noindent{\bf 2.6.1. Generalities}

In the regime $\lambda\ll\ell$ of interest,
the enhanced backscattering phenomenon
takes place in a narrow cone around the exact backscattering direction,
of angular width of order $\lambda/\ell$.
As recalled in the Introduction,
the shape of the cone of enhanced backscattering for a semi-infinite medium
is given by the sum of the cyclical, or maximally-crossed, diagrams.
This summation can be performed by means of an adaptation of RTT.
This property has been exploited extensively
in the case of scalar waves [6, 7, 19--21];
it has been extended more recently to polarisation effects
for electromagnetic waves [17, 18].

We restrict ourselves to a semi-infinite medium
and to normal incidence $(\theta_a=0)$.
We define the dimensionless transverse wavevector of the outgoing radiation as
$$
\Q=\q\ell
,\eqno(2.39)
$$
with a magnitude
$$
Q=q\ell=k\ell\theta=k_1\ell\theta_1
,\eqno(2.40)
$$
with $\theta_1$ being the observation angle.
We assume for definiteness that the vector $\Q$ is parallel to the $x$-axis,
namely $\Q=Q\vec{x}$, with $Q\ge0$.
In order to cure the ill-definedness of the co-ordinate system
at strictly normal incidence, we choose to give the initial wavevector
an infinitesimally small positive component along $\vec{x}$.
We thus set $\theta_a=0^+$, $\varphi_a=0$,
so that $\vec{\thetab}=\vec{x}$ and $\vec{\varphib}=\vec{y}$.
We then introduce a $Q$-dependent matrix of bistatic coefficients,
$\g_{ij}(Q,\mu_a,\varphi_a,\mu_b,\varphi_b)$.
The latter is defined, in analogy with eq. (2.16),
in terms of the matrix source function
$\G_{ij}(Q,\mu,\varphi,\mu_a,\varphi_a)$.
This matrix solves the $Q$-dependent SM equation,
obtained by replacing in eq. (2.13) the exponential damping factor
$\exp(-\tau/\mu')$ by $\exp\big(-(1-i\Q.\n)\tau/\mu'\big)$,
where $\n$ is the unit vector in the direction $(\theta',\varphi')$,
so that $\Q.\n=Q\nu'\cos\varphi'$.

We now turn to the explicit shape of the enhanced backscattering cone.
It can be expressed [17, 18] in terms of the values at normal incidence
of the bistatic coefficients,
$\g_{ij}(Q)=\g_{ij}(Q,\mu_a=1,\varphi_a=0,\mu_b=1,\varphi_b=0)$.
To be more specific, the total reflected intensity
near the backscattering direction,
i.e., for $\theta\ll 1$, $k\ell\gg 1$, and $Q=k\ell\theta\ge0$ fixed, reads
$$
A(Q)=A^L+A^C(Q)-A^{SS}
={4\over\pi(m+1)^4}\inner{\I_b}{\L.\bigl(\g^L+\g^C(Q)-\g^{SS}\bigr)}{\I_a}
.\eqno(2.41)
$$

\noindent$\bullet$
The first term in eq. (2.41), given by the sum of the ladder diagrams,
coincides with the expression (2.32) for the background reflected intensity.
At normal incidence it assumes the general form
$$
\g^L=\g(\mu_a=1,\varphi_a=0,\mu_b=1,\varphi_b=0)=\pmatrix{
\g_{11}&\g_{12}&0&0\cr\g_{12}&\g_{11}&0&0\cr
0&0&\g_{12}-\g_{11}&0\cr0&0&0&\g_{44}\cr
}
,\eqno(2.42)
$$
where the three constants $\g_{11}$, $\g_{12}$, and $\g_{44}$ only depend
on the index mismatch.

\noindent$\bullet$
The second term in eq. (2.41) is given by the sum of the maximally crossed,
or cyclical, diagrams.
It represents the contributions of the interference between the sequences
of any number $(N\ge 1)$ of scattering events
and their time-reversed counterparts.
At normal incidence we have [17, 18]
$$
\g^C(Q)=\pmatrix{
\g_{11}(Q)&\w\g_{12}(Q)&0&0\cr\w\g_{12}(Q)&\g_{22}(Q)&0&0\cr
0&0&\w\g_{33}(Q)&0\cr0&0&0&\w\g_{44}(Q)\cr}
,\eqno(2.43)
$$
with
$$
\eqalign{
&\w\g_{12}(Q)={1\over 2}\big(\g_{44}(Q)-\g_{33}(Q)\big),\cr
&\w\g_{33}(Q)={1\over 2}\big(\g_{33}(Q)+\g_{44}(Q)\big)-\g_{12}(Q),\cr
&\w\g_{44}(Q)={1\over 2}\big(\g_{33}(Q)+\g_{44}(Q)\big)+\g_{12}(Q).
}
\eqno(2.44)
$$

\noindent$\bullet$
The subtracted third term in eq. (2.41) is the contribution
of the single-scattering $(N=1)$ events,
which are their own time-reversed counterparts, and must not be double-counted.
At normal incidence it reads
$$
\g^{SS}=\frac{3}{4}\pmatrix{1&0&0&0\cr 0&1&0&0\cr 0&0&-1&0\cr 0&0&0&-1\cr}
.\eqno(2.45)
$$

The actual calculation of the $Q$-dependent bistatic matrix
$\g(Q,\mu_a,\varphi_a,\mu_b,\varphi_b)$ goes as follows.
By expanding the $Q$-dependent SM equation in the trigonometric polynomials
$\bigl\{e^{ik\varphi}\bigr\}$,
we obtain the following system of coupled equations $(-2\le k\le 2)$
$$
\eqalign{
&\G\i{k}(\tau,\mu)=\P\i{k}(\mu,\mu_a).\I_ae^{-ik\varphi_a-\tau/\mu_a}\cr
&+\int_0^\tau\!\d\tau'\!\int_0^1{\d\mu'\over2\mu'}e^{-(\tau-\tau')/\mu'}
\P\i{k}(\mu,\mu')
.\sum_{j=-2}^2 i^{k-j}J_{k-j}\big(Q(\tau-\tau')\nu'/\mu'\big)
\G\i{j}(\tau',\mu')\cr
&+\int_\tau^{\infty}\!\d\tau'\!\int_0^1{\d\mu'\over2\mu'}e^{-(\tau'-\tau)/\mu'}
\P\i{k}(\mu,-\mu')
.\sum_{j=-2}^2 i^{k-j}J_{k-j}\big(Q(\tau'-\tau)\nu'/\mu'\big)
\G\i{j}(\tau',-\mu')\cr
&+\int_0^{\infty}\!\d\tau'\!\int_0^1{\d\mu'\over2\mu'}e^{-(\tau+\tau')/\mu'}
\R(\mu').\P\i{k}(\mu,\mu')
.\sum_{j=-2}^2 i^{k-j}J_{k-j}\big(Q(\tau+\tau')\nu'/\mu'\big)
\G\i{j}(\tau',\mu'),\cr
}
\eqno(2.46)
$$
where $\nu$ has been defined in eq. (2.3),
and where the $J_n(z)$ are the Bessel functions, which admit the integral
representation
$$
i^nJ_n(z)=\int_0^{2\pi}{\d\varphi\over
2\pi}\exp\bigl(iz\cos\varphi-in\varphi\bigr)
,\eqno(2.47)
$$
and possess the symmetry property
$$
J_{-n}(z)=J_n(-z)=(-1)^nJ_n(z)
.\eqno(2.48)
$$
\smallskip
\noindent{\bf 2.6.2. Linear polarisations}

We now investigate the case where the initial beam is linearly polarised,
and a linear polarisation of the outgoing beam is detected.
Let $\psi_a$ and $\psi_b$ be the respective angles
between the directions of the polarisations and the direction
of the $\Q$-vector, i.e., the positive $x$-axis.
The corresponding Stokes vectors read
$$
\I_a=\pmatrix{\cos^2\psi_a\cr\sin^2\psi_a\cr\sin(2\psi_a)\cr0\cr},\quad
\I_b=\pmatrix{\cos^2\psi_b\cr\sin^2\psi_b\cr\sin(2\psi_b)\cr0\cr}
.\eqno(2.49)
$$
By inserting these expressions into the results (2.41--45),
we obtain that $A^L$ and $A^{SS}$ only depend on the relative angle
$$
\Psi=\psi_b-\psi_a
\eqno(2.50)
$$
between the directions of both polarisations, namely
$$
\eqalign{
A^L&={4\over\pi(m+1)^4}\bigl(\g_{11}\cos^2\Psi+\g_{12}\sin^2\Psi\bigr),\cr
A^{SS}&={3\over\pi(m+1)^4}\cos^2\Psi,
}\eqno(2.51)
$$
whereas $A^C(Q)$ depends separately on both polarisation directions:
$$
\eqalign{
A^C(Q)={4\over\pi(m+1)^4}
&\biggl[\g_{11}(Q)\cos^2\psi_a\cos^2\psi_b
+\g_{22}(Q)\sin^2\psi_a\sin^2\psi_b\cr
&+\bigl(2\g_{12}(Q)-\g_{33}(Q)-\g_{44}(Q)\bigr)
\cos\psi_a\sin\psi_a\cos\psi_b\sin\psi_b\cr
&+{1\over 2}\bigl(\g_{44}(Q)-\g_{33}(Q)\bigr)
\bigl(\sin^2\psi_a\cos^2\psi_b+\sin^2\psi_b\cos^2\psi_a\bigr)\biggr].
}
\eqno(2.52)
$$

We define as usual the enhancement factor $B(Q)$ as the ratio
between the total reflected intensity and its background value:
$$
B(Q)={A^L+A^C(Q)-A^{SS}\over A^L}
.\eqno(2.53)
$$

Right at the top of the backscattering cone,
corresponding to the exact backscattering direction $(Q=0)$,
the expressions (2.41), (2.43) simplify to
$$
A^C(0)={4\over\pi(m+1)^4}\Bigl(\g_{11}
-\frac{1}{2}(\g_{11}+\g_{12}-\g_{44})\sin^2\Psi\Bigr)
.\eqno(2.54)
$$
The enhancement factor thus reads
$$
B(0)={\left(2\g_{11}-\frd{3}{4}\right)
+\frd{1}{2}\left(-3\g_{11}+\g_{12}+\g_{44}+\frd{3}{2}\right)\sin^2\Psi
\over\g_{11}+\bigl(\g_{12}-\g_{11}\bigr)\sin^2\Psi}
.\eqno(2.55)
$$
The maximal enhancement factor $B_\l$ is observed for parallel detection, i.e.,
$\Psi=0$,
whereas the minimum $B_\r$ corresponds to perpendicular detection, i.e.,
$\Psi=\pm\pi/2$.
These extremal values read
$$
B_\l=2-{3\over 4\g_{11}},\quad
B_\r={\g_{11}+\g_{12}+\g_{44}\over 2\g_{12}}
.\eqno(2.56)
$$

A celebrated and universal feature of the enhanced backscattering cone
is the triangular shape of its top.
Within the present formalism, and in analogy with previous studies [19--21],
this phenomenon is described as follows.
For $Q\ll 1$, and for $i,j=1,2$,
the solution $\G\i{0}_{ij}(Q,\tau,\mu)$ of the $Q$-dependent SM equation
has a term linear in $Q$ that is proportional to the homogeneous solution
$(\GH)_i(\tau,\mu)$, namely
$$
\G_{ij}(Q,\tau,\mu)=\G_{ij}(\tau,\mu)-C_i\,Q(\GH)_i(\tau,\mu)+\ca{O}(Q^2)
\quad(i,j=1,2)
.\eqno(2.57)
$$
The constants $C_i$ are then fixed by requiring that
the above solution falls off as $\exp(-Q\tau)$ for $Q\tau\gg 1$.
This general property will be checked explicitly in section 3
in the absence of internal reflections.
We thus obtain $C_1=C_2=\tau_1(1)=\tau_2(1)$, so that
$$
\g_{ij}(Q)=\g_{ij}-\frac{2}{3}\tau_1(1)^2\,Q+\ca{O}(Q^2)\quad(i,j=1,2)
,\eqno(2.58)
$$
and finally
$$
A^C(Q)={4\over\pi(m+1)^4}\left(\g_{11}
-\frac{1}{2}\bigl(\g_{11}+\g_{22}-\g_{44}\bigr)\sin^2\Psi
-\frac{2}{3}\tau_1(1)^2\cos^2\Psi\,\,Q+\ca{O}(Q^2)\right)
.\eqno(2.59)
$$
Along the lines of refs. [19--21],
we define the width $\Delta Q$ of the triangular cone as
$$
A^C(Q)=A^C(0)\left(1-{Q\over\Delta Q}+\ca{O}(Q^2)\right)
.\eqno(2.60)
$$
The sharpest cone, namely the smallest width $\Delta Q$,
is observed for parallel detection, i.e., $\Psi=0$, where we have
$$
\Delta Q_\l={3\g_{11}\over 2\tau_1(1)^2}
.\eqno(2.61)
$$

The universal features of the top of the enhanced backscattering cone
described so far only depend on $Q$ and $\Psi$.
The full shape of the enhancement factor $B(Q)$ weakly depends
separately on the directions $\psi_a$ and $\psi_b$ of both polarisations.
This phenomenon will be illustrated in section 3.2
in the absence of internal reflections.
\smallskip
\noindent{\bf 2.6.3. Circular polarisations}

We end up by investigating the case of circularly polarised beams
at normal incidence.
The corresponding Stokes vectors now read
$$
\I_a=\displaystyle{\pmatrix{{1\over 2}\cr{1\over 2}\cr 0\cr\s_a\cr}},\quad
\I_b=\pmatrix{{1\over 2}\cr{1\over 2}\cr 0\cr\s_b}
,\eqno(2.62)
$$
where the helicity is $\s_a=1$ (respectively, $\s_a=-1$)
if the incident beam has a left (respectively, right) circular polarisation,
and similarly for the helicity $\s_b$ of the detection channel.
By inserting these expressions into the results (2.41--45),
we observe that the various backscattered amplitudes
only depend on the relative helicity
$$
\Sigma=\s_a\s_b
,\eqno(2.63)
$$
according to
$$
\eqalign{
A^{SS}&={3\over\pi(m+1)^4}(1-\Sigma),\cr
A^L&={2\over\pi(m+1)^4}(\g_{11}+\g_{12}+\g_{44}\Sigma),\cr
A^C(Q)&={1\over\pi(m+1)^4}\Bigl[\g_{11}(Q)+\g_{22}(Q)-\g_{33}(Q)+\g_{44}(Q)\cr
&{\hskip 60pt}+\bigl(2\g_{12}(Q)+\g_{33}(Q)+\g_{44}(Q)\bigr)\Sigma\Bigr].\cr
}
\eqno(2.64)
$$

The enhancement factor $B_{\Sigma}(Q)$, defined in analogy with eq. (2.53),
is larger for the helicity-preserving channel $(\Sigma=1)$
than in the channel of opposite helicity $(\Sigma=-1)$.
In particular, right at the top of the cone, we have
$$
B_1=2,\quad
B_{-1}={3\g_{11}-\g_{12}-\g_{44}-\frac{3}{2}\over\g_{11}+\g_{12}-\g_{44}}
.\eqno(2.65)
$$
The maximal value of the enhancement factor in the helicity-preserving channel
is exactly equal to two, because $A^L=A^C(0)$ and the single-scattering
contribution vanishes.
Corrections to this exact factor of two for denser diffusive media
$(\ell/\lambda$ not very large)
have been measured in a recent experiment [28],
and given a theoretical interpretation in terms of
recurrent double scattering [29].

Another consequence of the result (2.64)
is that the characteristic triangular shape of the cone only shows up in
the helicity-preserving channel.
The associated width, defined in analogy with eq. (2.60), reads
$$
\Delta Q_1={3(\g_{11}+\g_{12}+\g_{44})\over 4\tau_1(1)^2}
.\eqno(2.66)
$$
\smallskip
\noindent{\bf 3. EXACT SOLUTION IN THE ABSENCE OF INTERNAL REFLECTIONS}

This section is devoted to the exact
solution of the various SM equations introduced in section 2,
in the case where there is no optical index mismatch
between the sample and the surroundings,
so that there are no internal reflections:
the reflection matrix $\R(\mu)$ vanishes.
Therefore the SM equations (2.6), (2.12), (2.46)
involve convolution kernels,
which only depend on the difference of optical depths $\tau-\tau'$.
The problem is, however, still non-trivial
because of the semi-infinite geometry $(0<\tau<+\infty)$.
We have found it worthwhile to expose a self-contained derivation
of the Wiener-Hopf technique, and of the results known previously,
and already exposed in the book by Chandrasekhar [1].

The vector RTT problem is considered in section 3.1.
The outcomes concerning diffuse reflection and transmission
are compared in detail with those corresponding to multiple
isotropic scattering of scalar waves [19, 20].
Section 3.2 deals with the enhanced backscattering phenomenon.
We derive closed-form expressions for the five functions
describing the full shape of the enhanced backscattering cone,
up to the numerical solution of the $9\times 9$ system (3.74).
The present analysis thus goes one step further
than the recent work by Ozrin [18].
\smallskip
\noindent{\bf 3.1. Diffuse reflection and transmission}

In this section we derive the exact solution to the SM equations (2.6), (2.12)
in the absence of internal reflections,
obtaining thus predictions for the diffuse reflected and transmitted intensity.
We introduce the following parametrisation
$$
\eqalign{
&\G\i{0}(\tau,\mu)=\pmatrix{
A(\tau)+B(\tau)(1-\mu^2)\cr A(\tau)\cr 0\cr C(\tau)\mu\cr},\cr
&\G\i{1}(\tau,\mu)={\G\i{-1}}^*(\tau,\mu)=\nu\pmatrix{
D(\tau)\mu\cr 0\cr -iD(\tau)\cr E(\tau)\cr},\cr
&\G\i{2}(\tau,\mu)={\G\i{-2}}^*(\tau,\mu)=F(\tau)\pmatrix{
\mu^2\cr -1\cr -2i\mu\cr 0\cr},\cr
}
\eqno(3.1)
$$
where $\nu$ has been defined in eq. (2.3).
The functions $A(\tau),\cdots,F(\tau)$ obey the integral equations
$$
\eqalign{
&\left\{\matrix{
A=\frd{3}{4}\Bigl(\bigl(\mu_a^2\I_1+\I_2\bigr)e^{-\tau/\mu_a}+(M_0+M_2)\star
A+(M_2-M_4)\star B\Bigr),\hfill\cr
B=\frd{3}{4}\Bigl(\bigl((2-3\mu_a^2)\I_1-\I_2\bigr)e^{-\tau/\mu_a}
+(M_0-3M_2)\star A+(2M_0-5M_2+3M_4)\star B\Bigr),\hfill\cr
}\right.\cr
&{\hskip 12pt}C=\frac{3}{2}\Bigl(\mu_a\I_4e^{-\tau/\mu_a}+M_2\star C\Bigr),\cr
&{\hskip 12pt}D=\frac{3}{4}\Bigl(\nu_a(2\mu_a\I_1+i\I_3)e^{-\tau/\mu_a}
+(M_0+M_2-2M_4)\star D\Bigr),\cr
&{\hskip 12pt}E=\frac{3}{4}\Bigl(\nu_a\I_4e^{-\tau/\mu_a}+(M_0-M_2)\star
E\Bigr),\cr
&{\hskip 12pt}F=\frac{3}{8}\Bigl((\mu_a^2\I_1-\I_2+i\mu_a\I_3)e^{-\tau/\mu_a}
+(M_0+2M_2+M_4)\star F\Bigr),\cr
}
\eqno(3.2)
$$
where the brace shows that the equations for $A(\tau)$ and $B(\tau)$
are coupled, while the other four are decoupled.
In the above equations, the star denotes the convolution
between a kernel $M(\tau-\tau')$ and a function $A(\tau)$, defined as
$$
(M\star A)(\tau)=\int_0^{+\infty} M(\tau-\tau')A(\tau')\d\tau'
.\eqno(3.3)
$$
The kernels entering eq. (3.2) are the following even functions
$$
M_{2p}(\tau)=\int_0^1{\d\mu\over 2\mu}\mu^{2p}e^{-\vert\tau\vert/\mu}
.\eqno(3.4)
$$
\smallskip
\noindent{\bf 3.1.1. Preliminaries}

As recalled above, the integral equations (3.2) are exactly solvable
because of their convolution structure,
which suggests to utilise the Laplace transformation.
Along the lines of refs. [19--21],
the Laplace transform of a function $A(\tau)$ defined for $0<\tau<+\infty$
will be denoted by $a(s)$ (the corresponding lower-case letter), and defined as
$$
a(s)=\int_0^{+\infty} A(\tau)e^{s\tau}\d\tau
,\eqno(3.5)
$$
while the Laplace transform of the kernels $M_{2p}(\tau)$ read
$$
m_{2p}(s)=\int_{-\infty}^{+\infty} M_{2p}(\tau)e^{s\tau}\d\tau
=\int_0^1\d\mu{\mu^{2p}\over 1-s^2\mu^2}
,\eqno(3.6)
$$
i.e., explicitly,
$$
\eqalign{
&m_0(s)={1\over 2s}\ln{1+s\over 1-s},\cr
&m_2(s)={1\over s^2}\bigl(m_0(s)-1\bigr),\cr
&m_4(s)={1\over s^4}\left(m_0(s)-1-{s^2\over 3}\right).
}
\eqno(3.7)
$$
We also define for further use the following linear combinations of the
above kernels
$$
\eqalign{
&\phi_1(s)=1-\frac{3}{4}\bigl(m_0(s)-m_2(s)\bigr),\cr
&\phi_2(s)=-{1\over
s^2}\biggl[1-\frac{3}{2}\bigl(m_0(s)-m_2(s)\bigr)\biggr],\cr
&\phi_3(s)=1-\frac{3}{4}\bigl(m_0(s)+m_2(s)-2m_4(s)\bigr),\cr
&\phi_4(s)=1-\frac{3}{8}\bigl(m_0(s)+2m_2(s)+m_4(s)\bigr),\cr
&\phi_5(s)=1-\frac{3}{2}m_2(s),
}
\eqno(3.8)
$$
which we shall refer to as the kernel functions.
Both the kernels $m_{2p}(s)$ and the kernel functions $\phi_n(s)$
are even functions of $s$,
analytic in the $s$-plane cut along the real axis
from $-\infty$ to $-1$ and from $+1$ to $+\infty$.

In the following we shall need to factor the $\phi_n(s)$
into the corresponding so-called Wiener-Hopf $H$-functions,
defined after ref. [1] by the identity
$$
\phi_n(s)={1\over H_n(s)H_n(-s)}\quad(n=1,\cdots,5)
,\eqno(3.9)
$$
together with the condition that $H_n(s)$ is analytic
in the left half-plane $\re s<0$.
Consider first the case of a rational function of the form
$$
\phi(s)=\frd{\prod_{a=1}^M(s^2-z_a^2)}{\prod_{b=1}^N(s^2-p_b^2)}
,\eqno(3.10)
$$
with $2M$ zeros and $2N$ poles at arbitrary positions,
with $\re z_a>0$, $\re p_b>0$.
The factorisation (3.9) is elementary in this case, and the associated
$H$-function reads
$$
H(s)=\frd{\prod_{b=1}^N(s-p_b)}{\prod_{a=1}^M(s-z_a)}
.\eqno(3.11)
$$
This expression can be recast as a complex contour integral,
yielding thus an explicit representation of the $H$-functions in the general
case:
$$
H_n(s)=\exp\left(\int{\d z\over 2\pi i}\,{\phi'_n(z)\over\phi_n(z)}
\,\ln(z-s)\right)
=\exp\left(-\int{\d z\over 2\pi i}\,{\ln\phi_n(z)\over z-s}\,\right)
\quad(\re s<0)
,\eqno(3.12)
$$
where the vertical contour can be placed at $\re z=0$.
In the present case it is advantageous,
especially for the purpose of numerical evaluation,
to change variables from $z$ to an angle $\beta$ such that $z=i\tan\beta$.
We thus get
$$
H_n(s)=\exp\left({s\over\pi}\int_0^{\pi/2}
\d\beta{\ln\w\phi_n(\beta)\over\sin^2\beta+s^2\cos^2\beta}\right)
\quad(\re s<0)
,\eqno(3.13)
$$
with
$$
\eqalign{
\w\phi_1(\beta)&=1-\frac{3}{4}
\Bigl((\cot^2\beta+1)(\beta\cot\beta-1)+1\Bigr),\cr
\w\phi_2(\beta)&=\cot^2\beta\biggl[1-
\frac{3}{2}\Bigl((\cot^2\beta+1)(\beta\cot\beta-1)+1\Bigr)\biggr],\cr
\w\phi_3(\beta)&=1+
\frac{3}{4}\Bigl((2\cot^4\beta+\cot^2\beta-1)(\beta\cot\beta-1)+
\frac{2}{3}\cot^2\beta-1\Bigr),\cr
\w\phi_4(\beta)&=1-\frac{3}{8}\Bigl((\cot^2\beta-1)^2(\beta\cot\beta-1)+
\frac{1}{3}\cot^2\beta+1\Bigr),\cr
\w\phi_5(\beta)&=1+\frac{3}{2}\cot^2\beta(\beta\cot\beta-1).
}
\eqno(3.14)
$$

The following values of the $H$-functions will play a role hereafter.
First, the kernel functions have the following series expansions
around the origin
$$
\eqalign{
&\phi_1(s)=\frac{1}{2}-\frac{1}{10}s^2+\cdots,\quad
\phi_2(s)=\frac{1}{5}+\frac{3}{35}s^2+\cdots,\quad
\phi_3(s)=\frac{3}{10}-\frac{13}{70}s^2+\cdots,\cr
&\phi_4(s)=\frac{3}{10}-\frac{23}{70}s^2+\cdots,\quad
\phi_5(s)=\frac{1}{2}-\frac{3}{10}s^2+\cdots
}
\eqno(3.15)
$$
Eq. (3.9) yields $H_n(0)=1/\sqrt{\phi_n(0)}$, hence
$$
H_1(0)=\sqrt{2},\quad
H_2(0)=\sqrt{5},\quad
H_3(0)=H_4(0)=\sqrt{\frac{10}{3}},\quad
H_5(0)=\sqrt{2}
.\eqno(3.16)
$$
Second, for large $s$, namely $\vert s\vert\to+\infty$ with $\re s<0$,
the functions $H_n(s)$ with $n\ne 2$ go to unity, while we have
$H_2(s)\approx-s$.
Finally, the values of the $H$-functions at $s=-1$
can be accurately determined from the integral representation (3.13), (3.14).
We thus obtain
$$
\eqalign{
&H_1(-1)=1.277\,973,\quad H_2(-1)=3.469\,485,\quad H_3(-1)=1.465\,877,\cr
&H_4(-1)=1.396\,266,\quad H_5(-1)=1.203\,622.
}
\eqno(3.17)
$$
\smallskip
\noindent{\bf 3.1.2. Homogeneous SM equation and diffuse transmission}

The solution to the homogeneous SM equation is a priori of the form
$$
\GH(\tau,\mu)=\pmatrix{A_H(\tau)+B_H(\tau)(1-\mu^2)\cr A_H(\tau)\cr0\cr0\cr}
.\eqno(3.18)
$$
We deduce from the integral equations (3.2)
for the functions $A_H(\tau)$ and $B_H(\tau)$,
in the absence of source terms,
the following equations for their Laplace transforms $a_H(s)$ and $b_H(s)$
$$
\eqalign{
&\left(\frac{4}{3}
-m_0(s)-m_2(s)\right)a_H(s)+\bigl(m_4(s)-m_2(s)\bigr)b_H(s)=\ca{A}_H(s),\cr
&\bigl(3m_2(s)-m_0(s)\bigr)a_H(s)+\left(\frac{4}{3}
-2m_0(s)+5m_2(s)+3m_4(s)\right)b_H(s)=\ca{B}_H(s),\cr
}
\eqno(3.19)
$$
with right-hand sides
$$
\eqalign{
\ca{A}_H(s)&=\int{\d t\over 2\pi i(t-s)}
\Bigl[\bigl(m_0(t)+m_2(t)\bigr)a_H(t)+\bigl(m_2(t)-m_4(t)\bigr)b_H(t)\Bigr],\cr
\ca{B}_H(s)&=\int{\d t\over 2\pi i(t-s)}
\Bigl[\bigl(m_0(t)-3m_2(t)\bigr)a_H(t)+\bigl(2m_0(t)-5m_2(t)
-3m_4(t)\bigr)b_H(t)\Bigr].\cr
}
\eqno(3.20)
$$
On the other hand, the asymptotic behaviour (2.22) implies
$$
a_H(s)={1\over s^2}-{\tau_0\over s}+\ca{O}(1)\quad(s\to 0)
,\eqno(3.21)
$$
while $b_H(0)$ is expected to be finite in this limit.

The determinant of the $2\times 2$ linear system (3.19)
can be factorised as $(16/9)\phi_1(s)\phi_2(s)$.
This system can be put in diagonal form by looking for linear combinations
of the lines of eq. (3.19) involving only $\phi_1(s)$ or $\phi_2(s)$
acting on the unknowns.
We thus get
$$
\eqalignno{
&\phi_1(s)a_H(s)={3\over 8s^2}\Bigl((3-2s^2)\ca{A}_H(s)+\ca{B}_H(s)\Bigr),
&(3.22{\rm a})\cr
&\phi_2(s)\Bigl((1-s^2)b_H(s)-s^2 a_H(s)\Bigr)={3\over
4}\Bigl(\ca{A}_H(s)+\ca{B}_H(s)\Bigr).&(3.22{\rm b})
}
$$
We now solve these equations by means of the so-called
Wiener-Hopf technique.
We consider first eq. (3.22a), and we start by investigating
the case where $\phi_1(s)$ is a rational function of the form (3.10),
with zeros at $s=\pm z_{1,a}$ and poles at $s=\pm p_{1,b}$.
We observe that $a_H(s)$ is regular for $\re s<0$,
while the right-hand side of eq. (3.22a) is regular for $\re s>0$.
Moreover, eq. (3.20) implies that this right-hand side grows at most linearly
as
$s\to -\infty$.
Hence the solution of eq. (3.22a), normalised by the condition (3.21), reads
$$
a_H(s)={1-cs\over s^2}
\frd{\prod_{b=1}^N(1-s/p_{1,b})}{\prod_{a=1}^N(1-s/z_{1,a})}
={1-cs\over s^2}{H_1(s)\over\sqrt{2}}
,\eqno(3.23)
$$
where $c$ is a constant, yet to be determined.
Similarly, the solution of eq. (3.22b) reads
$$
b_H(s)={1\over 1-s^2}\left((1-cs){H_1(s)\over\sqrt{2}}-qH_2(s)\right)
,\eqno(3.24)
$$
where $q$ is another constant.
The notation $c$ and $q$ follows ref. [1].
These two constants are determined by expressing that the right-hand side
of eq. (3.24) remains finite as $s\to\pm 1$.
We finally obtain
$$
c={2H_2^2(-1)-H_1^2(-1)\over H_1^2(-1)+2H_2^2(-1)},\quad
q={2\sqrt{2}H_1(-1)H_2(-1)\over H_1^2(-1)+2H_2^2(-1)}
.\eqno(3.25)
$$
The representation (3.13), (3.14) permits a numerical evaluation of these
numbers,
and of all the subsequent quantities, with arbitrary accuracy.
We thus get
$$
c=0.872\,941,\quad q=0.487\,827
.\eqno(3.26)
$$
It is worth noticing that the exact solution derived above
does not require to determine the auxiliary functions $\ca{A}_H(s)$ and
$\ca{B}_H(s)$ explicitly.

The observables of interest can now be deduced as follows.

\noindent $\bullet$
The constant $\tau_0$ is obtained by comparing the result (3.23)
with the expansion (3.21), namely
$$
\tau_0=c-{H'_1(0)\over\sqrt{2}}=0.712\,110
.\eqno(3.27)
$$
This number is remarkably close to the celebrated value for
isotropic scattering of scalar waves, recalled in Table 2.

\noindent $\bullet$
The functions $\tau_1(\mu)$ and $\tau_2(\mu)$
are obtained from their definition (2.24), yielding
$$
\tau_1(\mu)=\frac{3}{2}\Bigl(a_H(-1/\mu)+(1-\mu^2)b_H(-1/\mu)\Bigr),\quad
\tau_2(\mu)=\frac{3}{2}a_H(-1/\mu)
,\eqno(3.28)
$$
i.e., explicitly,
$$
\tau_1(\mu)=\frac{3}{2}q\mu^2H_2(-1/\mu),\quad
\tau_2(\mu)=\frac{3}{2\sqrt{2}}\mu(\mu+c)H_1(-1/\mu)
.\eqno(3.29)
$$

In order to make a comparison with the case of scalar waves,
we must take into account
that the above results describe a single polarisation state,
and should be compared with {\it half} the
corresponding quantity for isotropic scattering of scalar waves,
determined in refs. [19, 20], and denoted there by $\tau_1(\mu)$,
and hereafter by $\tau\sca(\mu)$.
At nearly grazing incidence,
both functions $\tau_1(\mu)$ and $\tau_2(\mu)$
vanish linearly, according to
$$
\tau_1(\mu)\approx\frac{3}{2}q\mu=0.731\,740\,\mu,\quad
\tau_2(\mu)\approx{3\over 2\sqrt{2}}c\mu=0.925\,893\,\mu
,\eqno(3.30)
$$
while in the scalar case we have
$\tau\sca(\mu)/2\approx(\sqrt{3}/2)\mu=0.866\,025\,\mu$.
At normal incidence, both functions take the common value
$$
\tau_1(1)=\tau_2(1)={3\sqrt{2}H_1(-1)H_2^2(-1)\over H_1^2(-1)+2H_2^2(-1)}
=2.538\,761
,\eqno(3.31)
$$
which is again very close to the corresponding number
in the case of scalar waves (see Table 2).
The full functions $\tau_1(\mu)$ and $\tau_2(\mu)$ are plotted in Figure 1.
They hardly differ from each other,
and from half the corresponding scalar quantity $\tau\sca(\mu)/2$.

In order to underline the main novelty with respect to the scalar case [19,
20], namely polarisation effects, we plot in Figure 2 the degree
of polarisation $P$, defined in eq. (2.2), which reads in the present case
$$
P(\mu)={\tau_2(\mu)-\tau_1(\mu)\over\tau_2(\mu)+\tau_1(\mu)}
.\eqno(3.32)
$$
This quantity has a maximum at grazing incidence, namely
$$
P(0)={c-q\sqrt{2}\over c+q\sqrt{2}}=0.117\,127
,\eqno(3.33)
$$
and vanishes at normal incidence, as it should.
\smallskip
\noindent{\bf 3.1.3. Inhomogeneous SM equation and diffuse reflection}

The special solution of the full inhomogeneous SM equation
can be derived by solving the six equations (3.2),
along the lines of the previous subsection.

Let us begin with the functions $A(\tau)$ and $B(\tau)$.
Their Laplace transforms $a(s)$ and $b(s)$ still obey equations
of the form (3.19),
albeit with the contributions of the source terms
in their right-hand sides:
$$
\eqalign{
\ca{A}(s)&=\bigl(\mu_a^2\I_1+\I_2){\mu_a\over 1-s\mu_a}\cr
&+\int{\d t\over 2\pi i(t-s)}
\Bigl[\bigl(m_0(t)+m_2(t)\bigr)a(t)+\bigl(m_2(t)-m_4(t)\bigr)b(t)\Bigr],\cr
\ca{B}(s)&=\bigl((2-3\mu_a^2)\I_1-\I_2){\mu_a\over 1-s\mu_a}\cr
&+\int{\d t\over 2\pi i(t-s)}
\Bigl[\bigl(m_0(t)-3m_2(t)\bigr)a(t)+\bigl(2m_0(t)-5m_2(t)
-3m_4(t)\bigr)b(t)\Bigr].\cr
}
\eqno(3.34)
$$

These equations can be solved by means of the
Wiener-Hopf technique, along the lines of the previous subsection.
The undetermined constants can be fixed in terms of $c$ and $q$,
given by eq. (3.25), and we finally obtain
$$
\eqalign{
a(s)&=-{3\mu_aH_1(s)\over
2\sqrt{2}s}\left(q\mu_aH_2(-1/\mu_a)\I_1+{\mu_a+c-(1+\mu_ac)s\over
1-s\mu_a}{H_1(-1/\mu_a)\over\sqrt{2}}\I_2\right),\cr
b(s)&={1\over 1-s^2}\biggl[s^2a(s)\cr
&{\hskip 43pt}-{3\mu_aH_2(s)\over 2}\left(\mu_a{\mu_a-c-(1-\mu_ac)s\over
1-s\mu_a}H_2(-1/\mu_a)\I_1+q{H_1(-1/\mu_a)\over\sqrt{2}}\I_2\right)\biggr].
}
\eqno(3.35)
$$

The other four functions $C(\tau),\cdots,F(\tau)$ are easier to determine,
since the last four lines of eq. (3.2) are uncoupled.
We thus obtain the following closed-form expressions for their Laplace
transforms
$$
\eqalign{
c(s)&=\frac{3}{2}{\mu_a^2\I_4\over 1-s\mu_a}H_5(s)H_5(-1/\mu_a),\cr
d(s)&=\frac{3}{4}{\mu_a\nu_a(2\mu_a\I_1+i\I_3)\over
1-s\mu_a}H_3(s)H_3(-1/\mu_a),\cr
e(s)&=\frac{3}{4}{\mu_a\nu_a\I_4\over 1-s\mu_a}H_1(s)H_1(-1/\mu_a),\cr
f(s)&=\frac{3}{8}{\mu_a(\mu_a^2\I_1-\I_2+i\mu_a\I_3)\over
1-s\mu_a}H_4(s)H_4(-1/\mu_a).\cr
}
\eqno(3.36)
$$

The above results (3.35), (3.36) allow us to give the following expression
for the full bistatic matrix in the absence of internal reflections:
$$
\g(\mu_a,\varphi_a,\mu_b,\varphi_b)
=\sum_{k=-2}^2\g\i{k}(\mu_a,\mu_b)e^{ik(\varphi_a-\varphi_b)}
,\eqno(3.37)
$$
with
$$
\eqalign{
\g&\i{0}(\mu_a,\mu_b)=\frac{3}{2}\mu_a\mu_b\times\cr
&\times\pmatrix{
\mu_a\mu_b\left(\frd{1+\mu_a\mu_b}{\mu_a+\mu_b}-c\right)H_2^aH_2^b
&\frd{q\mu_b}{\sqrt{2}}H_1^aH_2^b&0&0\cr
\frd{q\mu_a}{\sqrt{2}}H_2^aH_1^b
&\frac{1}{2}\left(\frd{1+\mu_a\mu_b}{\mu_a+\mu_b}+c\right)H_1^aH_1^b&0&0\cr
0&0&0&0\cr
0&0&0&-\frd{\mu_a\mu_b}{\mu_a+\mu_b}H_5^aH_5^b\cr},\cr\cr
\g&\i{1}(\mu_a,\mu_b)={\g\i{-1}}^*(\mu_a,\mu_b)
=\frac{3}{4}{\mu_a\nu_a\mu_b\nu_b\over\mu_a+\mu_b}
\pmatrix{-2\mu_a\mu_bH_3^aH_3^b&0&-i\mu_bH_3^aH_3^b&0\cr0&0&0&0\cr
-2i\mu_aH_3^aH_3^b&0&H_3^aH_3^b&0\cr
0&0&0&H_1^aH_1^b},\cr\cr
\g&\i{2}(\mu_a,\mu_b)={\g\i{-2}}^*(\mu_a,\mu_b)
=\frac{3}{8}{\mu_a\mu_b\over\mu_a+\mu_b}\pmatrix{
\mu_a^2\mu_b^2&-\mu_b^2&i\mu_a\mu_b^2&0\cr-\mu_a^2&1&-i\mu_a&0\cr
2i\mu_a^2\mu_b&-2i\mu_b&-2\mu_a\mu_b&0\cr0&0&0&0\cr
}H_4^aH_4^b,\cr
}
\eqno(3.38)
$$
and with the notation $H_n^a=H_n(-1/\mu_a)$, $H_n^b=H_n(-1/\mu_b)$.

At normal incidence $(\mu_a=\mu_b=1,\;\varphi_a=\varphi_b=0)$,
the above expression simplifies, and it agrees with the general form (2.42),
with
$$
\eqalign{
&\g_{11}={3H_1^2(-1)H_2^2(-1)\over H_1^2(-1)+2H_2^2(-1)}+{3H_4^2(-1)\over
8}=3.025\,270,\cr
&\g_{12}={3H_1^2(-1)H_2^2(-1)\over H_1^2(-1)+2H_2^2(-1)}-{3H_4^2(-1)\over
8}=1.563\,100,\cr
&\g_{44}=-{3H_5^2(-1)\over 4}=-1.086\,530.\cr
}
\eqno(3.39)
$$

We now derive a few special results of interest
from the above general expressions.
First, neglecting polarisation effects,
the total diffuse reflected intensity in the normal direction is given by
$$
\g_{11}+\g_{12}={6H_1^2(-1)H_2^2(-1)\over H_1^2(-1)+2H_2^2(-1)}=4.588\,369
.\eqno(3.40)
$$
This number is slightly above the corresponding value $\gamma(1,1)$
for isotropic scattering of scalar waves (see Table 2).

The interesting polarisation dependence of the enhancement factor
at the top of the backscattering cone, described in general terms in section
2.6, can also be made fully quantitative in the present case.
For linearly polarised beams, the extremal enhancement factors (2.56),
corresponding to parallel and perpendicular detection, read
$$
B_\l=1.752\,088,\quad B_\r=1.120\,158
,\eqno(3.41)
$$
while the width of the triangular cone for parallel detection (2.61) is
$$
\Delta Q_\l=0.704\,063
.\eqno(3.42)
$$
For circularly polarised beams,
the enhancement factors (2.65) in the helicity-preserving channel
and in the channel of opposite helicity read, respectively,
$$
B_1=2,\quad B_{-1}=1.250\,989
,\eqno(3.43)
$$
while the width of the triangular cone in the helicity-preserving channel
(2.66) is
$$
\Delta Q_1=0.407\,487
.\eqno(3.44)
$$
The most significant of these numbers are again compared with their analogues
for isotropic scattering of scalar waves in Table 2.
\smallskip
\noindent{\bf 3.1.4. Extinction lengths of non-diffusive modes}

The exact solution of the inhomogeneous SM equation
in the absence of internal reflections,
derived in section 3.1.3, also allows us to predict the extinction lengths
characterising the exponential fall-off
of the various non-diffusive polarised components
of the intensity of radiation, deep in the bulk of a semi-infinite sample.
These quantities do not depend at all on the index mismatch,
so that the results derived below are quite general.

Let us take for definiteness the example of the component of the intensity
described by the function $D(\tau)$, defined in eq. (3.1).
Its Laplace transform $d(s)$ is by construction analytic for $\re s<0$.
The explicit expression (3.36) shows, however, that it is actually
analytic in a larger domain, defined by $\re s<1/\mu_a$ and $\re s<s_3$,
where $s_3$ is the first pole of $H_3(s)$,
namely the first zero of the kernel function $\phi_3(s)$.
Here {\it first} means {\it having the smallest real part}.
We have $s_3=0.914\,815<1\le 1/\mu_a$.
The first singularity of $d(s)$ is therefore a simple pole at $s=s_3$.
We have thus demonstrated the exponential fall-off
$D(\tau)\sim\exp(-\tau/\ell_3)$,
with a dimensionless reduced extinction length $\ell_3=1/s_3=1.093\,116$.

More generally, all the extinction lengths are given by the locations
of the first singularities of the corresponding Laplace transforms.
We thus obtain
$$
B(\tau)\sim E(\tau)\sim e^{-\tau/\ell_1},\quad
C(\tau)\sim e^{-\tau/\ell_5},\quad
D(\tau)\sim e^{-\tau/\ell_3},\quad
F(\tau)\sim e^{-\tau/\ell_4}
,\eqno(3.45)
$$
while the function $A(\tau)$, pertaining to the diffusive sector,
does not fall off, but it rather admits the limit value
$A(+\infty)=\tau_1(\mu_a)\I_1+\tau_2(\mu_a)\I_2$ deep inside a semi-infinite
sample.

Thus there are altogether four different
dimensionless extinction lengths, $\ell_n=1/s_n$ $(n=1, 3, 4, 5)$,
where $s_n$ is the zero of the kernel function $\phi_n(s)$
having the smallest positive real part.
Table 3 gives the exact values of the extinction lengths $\ell_n$,
together with the corresponding approximate values $\ell_n\dif$,
predicted by the diffusion approximation [10].
This approximate scheme consists in brutally truncating
the kernel functions $\phi_n(s)$
to the second order of their series expansion in $s$, given in eq. (3.15).
\smallskip
\noindent{\bf 3.2. Enhanced backscattering peak}

We now derive analytical expressions for the five functions
describing the polarisation dependence of the enhanced backscattering cone
at normal incidence,
according to the general formalism exposed in section 2.6,
in the absence of internal reflections.
We start by introducing the following parametrisation
$$
\eqalign{
&\G\i{0}(\tau,\mu)=\pmatrix{
A(\tau)+B(\tau)(1-\mu^2)\cr A(\tau)\cr 0\cr C(\tau)\mu\cr},\cr
&\G\i{1}(\tau,\mu)=-{\G\i{-1}}^*(\tau,\mu)=\nu\pmatrix{
\bigl(iD(\tau)-E(\tau)\bigr)\mu\cr 0\cr D(\tau)+iE(\tau)\cr iF(\tau)\cr},\cr
&\G\i{2}(\tau,\mu)={\G\i{-2}}^*(\tau,\mu)=\bigl(G(\tau)+iH(\tau)\bigr)\pmatrix{
\mu^2\cr -1\cr -2i\mu\cr 0\cr},\cr
}
\eqno(3.46)
$$
that slightly differs from eq. (3.1).

Eq. (2.46) implies that the
eight real functions $A(\tau),\cdots,H(\tau)$ obey the following
sets of coupled linear integral equations, as shown by the braces
$$
\eqalignno{
&\left\{\matrix{
A=\frd{3}{4}\Bigl((\I_1+\I_2)e^{-\tau}+(M_{00}+M_{02})\star A+(M_{02}-M_{04})
\star B\hfill\cr
{\hskip 32pt}-2M_{13}\star D+2(M_{20}+M_{22})\star G\Bigr),\hfill\cr\cr
B=\frd{3}{4}\Bigl(-(\I_1+\I_2)e^{-\tau}+(M_{00}-3M_{02})\star A
+(2M_{00}-5M_{02}+3M_{04})\star B\hfill\cr
{\hskip 32pt}+2(-2M_{11}+3M_{13})\star D-2(M_{20}+3M_{22})\star G\Bigr),
\hfill\cr
D=\frd{3}{4}\Bigl(2M_{11}\star A+2(M_{11}-M_{13})\star B\hfill\cr
{\hskip 32pt}+(M_{00}+M_{02}-2M_{04}+M_{20}-2M_{22})\star
D+2(M_{11}+M_{13}+M_{31})\star G\Bigr),\hfill\cr
G=\frd{3}{8}\Bigl((\I_1-\I_2)e^{-\tau}+M_{20}\star A-M_{22}\star B\hfill\cr
{\hskip 32pt}-(M_{11}+M_{13}+M_{31})\star D+(M_{00}+2M_{02}+M_{04}+M_{40})\star
G\Bigr),\hfill\cr
}\right.&(3.47{\rm a})\cr\cr
&\left\{\matrix{
C=\frd{3}{2}\Bigl(\I_4e^{-\tau}+M_{02}\star C-2M_{11}\star F\Bigr),\hfill\cr\cr
F=\frd{3}{4}\Bigl(M_{11}\star C+(M_{00}-M_{02}-M_{20})\star F\Bigr),\hfill\cr
}\right.&(3.47{\rm b})\cr\cr
&\left\{\matrix{
E=\frd{3}{4}\Bigl((M_{00}+M_{02}-2M_{04}-M_{20}+2M_{22})\star
E+2(M_{11}+M_{13}-M_{31})\star H\Bigr),\hfill\cr\cr
H=\frd{3}{8}\Bigl(\I_3e^{-\tau}+(-M_{11}-M_{13}+M_{31})\star
E+(M_{00}+2M_{02}+M_{04}-M_{40})\star H\Bigr).\hfill\cr
}\right.&(3.47{\rm c})\cr
}
$$
The kernels involved in these equations read
$$
M_{pq}(Q,\tau)=\int_0^1{\d\mu\over 2\mu}\mu^p\nu^qe^{-\vert\tau\vert/\mu}
J_q\bigl(Q\nu\vert\tau\vert/\mu\bigr)
,\eqno(3.48)
$$
where $\nu$ has been defined in eq. (2.3).
\smallskip
\noindent{\bf 3.2.1. Preliminaries: kernels and $H$-functions}

The explicit solution of eq. (3.47) again involves the Laplace transforms
$a(s),\cdots,h(s)$ of the functions introduced in the parametrisation (3.46).
In a first step, we must therefore
evaluate the Laplace transforms $m_{pq}(Q,s)$ of the kernels $M_{pq}(Q,\tau)$.
Let us take the example of $m_{00}(Q,s)$.
The representation (3.48) yields
$$
m_{00}(Q,s)=\underbrace{\int_{-1}^1{\d\mu\over 2}\int_0^{2\pi}{\d\varphi\over
2\pi}}_{\displaystyle{\int{\d\Omega\over 4\pi}}}
\int_{-\infty}^{+\infty}{\d\tau\over\vert\mu\vert}
\Theta(\tau\mu)\exp\left(s\tau-{\tau\over\mu}
+iQ{\nu\tau\over\mu}\sin\varphi\right)
,\eqno(3.49)
$$
where $\Theta$ is Heaviside's function,
and $\d\Omega$ is the solid-angle element on the unit sphere of $\bf n$, with
co-ordinates $X=\nu\cos\varphi=\sin\theta\cos\varphi$,
$Y=\nu\sin\varphi=\sin\theta\sin\varphi$, $Z=\mu=\cos\theta$.
Performing the $\tau$-integral yields
$$
m_{00}(Q,s)=\int{\d\Omega\over 4\pi}{1\over
1-s\cos\theta-iQ\sin\theta\cos\varphi}
.\eqno(3.50)
$$
The denominator can be transformed by a suitable rotation
in the $Y$-$Z$ plane into $1-\s Z'$, with
$$
\s=\sqrt{s^2-Q^2}
.\eqno(3.51)
$$
We thus obtain
$$
m_{00}(Q,s)=m_0(\s)
.\eqno(3.52)
$$
It turns out that all the kernels involved in eq. (3.47)
can be treated similarly, and expressed as linear combinations of
the $m_{2p}(\s)$, which have been determined in eq. (3.7), namely
$$
\eqalign{
m_{00}(Q,s)&=m_0(\s\bigr),\cr
m_{02}(Q,s)&=m_2(\s)+{Q^2\over 2\s^2}\bigl(3m_2(\s)-m_0(\s)\bigr),\cr
m_{04}(Q,s)&=m_4(\s)+{Q^2\over\s^2}\bigl(5m_4(\s)-3m_2(\s)\bigr)+{Q^4\over
8\s^4}\bigl(35m_4(\s)-30m_2(\s)+3m_0(\s)\bigr),\cr
m_{11}(Q,s)&={Qs\over 2\s^2}\bigl(3m_2(\s)-m_0(\s)\bigr),\cr
m_{13}(Q,s)&={Qs\over 2\s^2}\bigl(5m_4(\s)-3m_2(\s)\bigr)+{Q^3s\over
8\s^4}\bigl(35m_4(\s)-30m_2(\s)+3m_0(\s)\bigr),\cr
m_{20}(Q,s)&={Q^2\over 2\s^2}\bigl(3m_2(\s)-m_0(\s)\bigr),\cr
m_{22}(Q,s)&={Q^2\over 4\s^2}\bigl(15m_4(\s)-12m_2(\s)+m_0(\s)\bigr)+{Q^4\over
8\s^4}\bigl(35m_4(\s)-30m_2(\s)+3m_0(\s)\bigr),\cr
m_{31}(Q,s)&={Q^3s\over 8\s^4}\bigl(35m_4(\s)-30m_2(\s)+3m_0(\s)\bigr),\cr
m_{40}(Q,s)&={Q^4\over 8\s^4}\bigl(35m_4(\s)-30m_2(\s)+3m_0(\s)\bigr).\cr
}
\eqno(3.53)
$$

We are again led to consider the kernel functions $\phi_n(\s)$,
defined in eq (3.8).
Although the latter only depend on the variable $\s$ of eq. (3.51),
the associated $H$-functions $H_n(Q,s)$ depend separately on both variables $Q$
and $s$.
The latter functions are still defined by the Wiener-Hopf identity
(3.9), which now reads
$$
\phi_n(\s)={1\over H_n(Q,s)H_n(Q,-s)}
,\eqno(3.54)
$$
with the condition that the $H_n(Q,s)$ be analytic in the left half-plane $\re
s<0$.
For a rational kernel function $\phi(\s)$ of the form (3.10), the $H$-function
reads
$$
H(Q,s)=\frd{\prod_{b=1}^N\left(s-\sqrt{Q^2+p_b^2}\right)}
{\prod_{a=1}^M\left(s-\sqrt{Q^2+z_a^2}\right)}
.\eqno(3.55)
$$
In the present case, the functions $H_n(Q,s)$ still possess the integral
representation (3.13), up to the replacement
$$
\beta\to\w\beta(Q,\beta)=\arctan\sqrt{Q^2+\tan^2\beta}
.\eqno(3.56)
$$
\smallskip
\noindent{\bf 3.2.2. Derivation of $\g_{44}(Q)$}

We begin with the exact solution of the $2\times 2$ system (3.47b),
yielding, after a Laplace transformation,
$c(Q,s)$ and $f(Q,s)$, and $\g_{44}(Q)$ through
$c(Q,-1)=-\g_{44}(Q)\I_4$.
The determinant of this linear system can be factorised as
$-(9/8)\phi_1(\s)\phi_5(\s)$.
Along the lines of section 3.1, the system is made diagonal by setting
$$
\eqalign{
\w c(Q,s)&=sc(Q,s)-2Qf(Q,s),\cr
\w f(Q,s)&=(Q/2)c(Q,s)-sf(Q,s).
}
\eqno(3.57)
$$
The inverse formulae read
$$
\eqalign{
\s^2c(Q,s)&=s\w c(Q,s)-2Q\w f(Q,s),\cr
\s^2f(Q,s)&=(Q/2)\w c(Q,s)-s\w f(Q,s),
}
\eqno(3.58)
$$
and the new functions obey
$$
\eqalign{
\phi_5(\s)\w c(Q,s)&=-{s\over 1-s}\I_4+\w\ca{C}(Q,s),\cr
\phi_1(\s)\w f(Q,s)&={Q\over 1-s}\I_4+\w\ca{F}(Q,s),
}
\eqno(3.59)
$$
where $\w\ca{C}(Q,s)$ and $\w\ca{F}(Q,s)$ are defined in analogy with eq.
(3.20).

The solution of eq. (3.59) reads
$$
\eqalign{
\w c(Q,s)&=\left(c_1+{c_2\over 1-s}\right)\I_4H_5(Q,s),\cr
\w f(Q,s)&=\left(f_1+{f_2\over 1-s}\right)\I_4H_1(Q,s).
}
\eqno(3.60)
$$
This expression involves, besides the corresponding $H$-functions,
four constants, yet to be determined.
The $s\to 1$ limit of eq. (3.59) fixes two of them:
$$
c_2={3\over2}\I_4H_5(Q,-1),\quad
f_2={3\over4}Q\I_4H_1(Q,-1)
.\eqno(3.61)
$$
The last two constants, $c_1$ and $f_1$,
are then fixed by requiring that $c(Q,s)$ and $f(Q,s)$,
as given by eq. (3.58), are finite for $\s\to 0$, i.e., $s\to Q$ and $s\to -Q$.
Skipping lengthy details, we finally get
$$
\g_{44}(Q)=-{3\over
4}{\ca{N}_{44}(Q)\over(1-Q^2)^2\bigl(H_1^2(Q,-Q)+H_5^2(Q,-Q)\bigr)}
,\eqno(3.62)
$$
where
$$
\eqalign{
\ca{N}_{44}(Q)
&=Q^2(1+Q)^2H_5^2(Q,-Q)H_1^2(Q,-1)+Q^2(1-Q)^2H_1^2(Q,-Q)H_1^2(Q,-1)\cr
&+(1-Q)^2H_5^2(Q,-Q)H_5^2(Q,-1)+(1+Q)^2H_1^2(Q,-Q)H_5^2(Q,-1)\cr
&-8Q^2H_1(Q,-Q)H_5(Q,-Q)H_1(Q,-1)H_5(Q,-1).
}
\eqno(3.63)
$$
\smallskip
\noindent{\bf 3.2.3. Derivation of $\g_{33}(Q)$}

The $2\times 2$ linear system (3.47c) yields,
after a Laplace transformation, $e(Q,s)$ and $h(Q,s)$,
and $\g_{33}(Q)$ through $h(Q,-1)=(1/4)\g_{33}(Q)\I_3$.
The determinant of this system can be factorised as
$(9/32)\phi_3(\s)\phi_4(\s)$.
Its exact solution closely follows the lines of the previous subsection.
We are led to consider the linear combinations
$$
\eqalign{
\w e(Q,s)&=-se(Q,s)+2Qh(Q,s),\cr
\w h(Q,s)&=-(Q/2)e(Q,s)+sh(Q,s),
}
\eqno(3.64)
$$
which obey
$$
\eqalign{
\phi_3(\s)\w e(Q,s)&=-{Q\over 1-s}\I_3+\ca{E}(Q,s),\cr
\phi_4(\s)\w h(Q,s)&=-{s\over 1-s}\I_3+\ca{H}(Q,s).
}
\eqno(3.65)
$$
The solution of these equations is fully similar to that of eq. (3.59).
We are thus left with
$$
\g_{33}(Q)=-{3\over
4}{\ca{N}_{33}(Q)\over(1-Q^2)^2\bigl(H_3^2(Q,-Q)+H_4^2(Q,-Q)\bigr)}
,\eqno(3.66)
$$
and with
$$
\eqalign{
\ca{N}_{33}(Q)
&=Q^2(1+Q)^2H_4^2(Q,-Q)H_3^2(Q,-1)+Q^2(1-Q)^2H_3^2(Q,-Q)H_3^2(Q,-1)\cr
&+(1-Q)^2H_4^2(Q,-Q)H_4^2(Q,-1)+(1+Q)^2H_3^2(Q,-Q)H_4^2(Q,-1)\cr
&-8Q^2H_3(Q,-Q)H_4(Q,-Q)H_3(Q,-1)H_4(Q,-1).
}
\eqno(3.67)
$$
\smallskip
\noindent{\bf 3.2.4. Derivation of $\g_{11}(Q)$, $\g_{12}(Q)$, and
$\g_{22}(Q)$}

The determinant of the $4\times 4$ linear system (3.47a),
after a Laplace transformation,
can be factorised as $(81/256)\phi_1(\s)\phi_2(\s)\phi_3(\s)\phi_4(\s)$.
Its exact solution, along the lines of the previous cases,
involves algebraic manipulations on very lengthy expressions.
Some of them have been either carried out, or just checked,
by means of the MACSYMA software.
Furthermore, the final step of this calculation,
involving the solution of the $9\times 9$ linear system (3.74),
must for practical purposes be performed numerically.

In a first step, the system (3.47a) is put in diagonal form
by introducing the linear combinations
$$
\eqalign{
\w a(Q,s)&=2\s^2a(Q,s)-Q^2b(Q,s)-2Qsd(Q,s)+2Q^2g(Q,s),\cr
\w b(Q,s)&=-2\s^4a(Q,s)+(-2\s^4-2Q^2\s^2+2\s^2+3Q^2)b(Q,s)\cr
&+2Qs(-2\s^2+3)d(Q,s)+2Q^2(2\s^2-3)g(Q,s),\cr
\w d(Q,s)&=Qsb(Q,s)+(\s^2+2Q^2)d(Q,s)-2Qsg(Q,s),\cr
\w g(Q,s)&=(Q^2/2)b(Q,s)+Qsd(Q,s)-(2\s^2+Q^2)g(Q,s).\cr
}
\eqno(3.68)
$$
These new functions obey
$$
\eqalign{
\phi_1(\s)\w a(Q,s)&={2(s^2-1)(\I_1+\I_2)-2Q^2\I_2\over 1-s}+\w\ca{A}(Q,s),\cr
\phi_2(\s)\w b(Q,s)&={2Q^2\I_1\over 1-s}+\w\ca{B}(Q,s),\cr
\phi_3(\s)\w d(Q,s)&={2Qs\I_1\over 1-s}+\w\ca{D}(Q,s),\cr
\phi_4(\s)\w g(Q,s)&={2s^2(\I_1-\I_2)+2Q^2\I_2\over 1-s}+\w\ca{G}(Q,s),\cr
}
\eqno(3.69)
$$
where $\w\ca{A}(Q,s)$, $\w\ca{B}(Q,s)$, $\w\ca{D}(Q,s)$,
and $\w\ca{G}(Q,s)$ are defined in analogy with eq. (3.20).

The formulae inverse to eq. (3.68) read
$$
\eqalign{
4\s^4(\s^2-1)a(Q,s)&=\s^2(2\s^2+Q^2-2)\w a(Q,s)+Q^2\w b(Q,s)\cr
&+4Qs(\s^2-1)\w d(Q,s)-2Q^2(\s^2-1)\w g(Q,s),\cr
4\s^4(\s^2-1)b(Q,s)&=-\s^2(2\s^2+3Q^2)\w a(Q,s)-(2\s^2+3Q^2)\w b(Q,s)\cr
&-12Qs(\s^2-1)\w d(Q,s)+6Q^2(\s^2-1)\w g(Q,s),\cr
2\s^4(\s^2-1)d(Q,s)&=Qs\s^2\w a(Q,s)+Qs\w b(Q,s)\cr
&+2(\s^2-1)(\s^2+2Q^2)\w d(Q,s)-2Qs(\s^2-1)\w g(Q,s),\cr
8\s^4(\s^2-1)g(Q,s)&=Q^2\s^2\w a(Q,s)+Q^2\w b(Q,s)\cr
&+4Qs(\s^2-1)\w d(Q,s)-2(\s^2-1)(2\s^2+Q^2)\w g(Q,s),
}
\eqno(3.70)
$$
so that we have
$$
\eqalign{
\g_{11}(Q)\I_1+\g_{12}(Q)\I_2&=-{\w b(Q,-1)+4Q\w d(Q,-1)+2\w
g(Q,-1)\over 2(1-Q^2)^2},\cr
\g_{21}(Q)\I_1+\g_{22}(Q)\I_2&={\w a(Q,-1)+2\w g(Q,-1)\over 2(1-Q^2)}.
}
\eqno(3.71)
$$

The solution to eq. (3.69) reads
$$
\eqalign{
\w a(Q,s)&=\left(a_1+a_2s+{a_3\over 1-s}\right)H_1(Q,s),\cr
\w b(Q,s)&=\left(b_1+b_2s+b_3s^2+{b_4\over 1-s}\right)H_2(Q,s),\cr
\w d(Q,s)&=\left(d_1+d_2s+{d_3\over 1-s}\right)H_3(Q,s),\cr
\w g(Q,s)&=\left(g_1+g_2s+{g_3\over 1-s}\right)H_4(Q,s),\cr
}
\eqno(3.72)
$$
where $a_1,\cdots,g_3$ are 13 $Q$-dependent constants to be determined.

The $s\to 1$ limit of eq. (3.69) fixes four of these constants:
$$
\matrix{
\displaystyle a_3=-{3\over2}Q^2\I_2H_1(Q,-1),\hfill
&b_4=-3Q^2\I_1H_2(Q,-1),\hfill\cr\cr
\displaystyle d_3=-{3\over2}Q\I_1H_3(Q,-1),\hfill
&\displaystyle g_3=-{3\over4}\bigl(\I_1+(Q^2-1)\I_2\bigr)H_4(Q,-1).\hfill\cr
}
\eqno(3.73)
$$
The last nine constants are then determined
by expressing that the functions $a(Q,s)$, $\cdots$, $g(Q,s)$
have the expected regularity properties
at the points where the inversion formulas (3.70) are singular,
namely $\sigma^2=0$, i.e., $s=\pm Q$, or $\sigma^2=1$,
i.e., $s=\pm\sqrt{1+Q^2}$.
We thus obtain the following system of nine linear equations
$$
\eqalignno{
&\w a\left(Q,-\sqrt{1+Q^2}\right)+\w b\left(Q,-\sqrt{1+Q^2}\right)=0,&(3.74{\rm
a})\cr
&\w a\left(Q,\sqrt{1+Q^2}\right)+\w b\left(Q,\sqrt{1+Q^2}\right)=0,&(3.74{\rm
b})\cr
&\w a(Q,-Q)+2\w g(Q,-Q)=0,&(3.74{\rm c})\cr
&\w b(Q,-Q)-6\w g(Q,-Q)=0,&(3.74{\rm d})\cr
&\w d(Q,-Q)+2\w g(Q,-Q)=0,&(3.74{\rm e})\cr
&{\d\over\d s}\Bigl(\w b(Q,s)+4\w d(Q,s)+2\w g(Q,s)\Bigr)_{s=-Q}-8Q\w
g(Q,-Q)=0,&(3.74{\rm f})\cr
&\w b(Q,Q)-6\w g(Q,Q)=0,&(3.74{\rm g})\cr
&\w d(Q,Q)-2\w g(Q,Q)=0,&(3.74{\rm h})\cr
&3{\d\over\d s}\Bigl(\w b(Q,s)-4\w d(Q,s)+2\w g(Q,s)\Bigr)_{s=Q}+10Q\w
a(Q,Q)+44Q\w g(Q,Q)=0,&(3.74{\rm i})\cr
}
$$
where
(a) and (b) express the regularity of the functions $a(Q,s),\cdots,g(Q,s)$
(absence of pole) at $s=\pm\sqrt{1+Q^2}$;
(c) to (f) express their regularity (absence of double and of simple pole)
at $s=-Q$;
(g) to (i) express the absence of double pole at $s=Q$,
as well as the proportionality of the residues of the simple poles to
the right null vector of the system (3.47a),
${\bf V}=(3-2Q^2,3+Q^2,2Q^2,Q^2/2)$.

The last of the above properties
implies that the four functions $A(\tau)$, $B(\tau)$, $D(\tau)$, and $G(\tau)$
fall off as $\exp(-Q\tau)$.
The dimensionless extinction length in the diffusive sector
thus reads $\ell(Q)=1/Q$ in units of the mean free path $\ell$, i.e.,
$$
L(q)={\ell\over Q}={1\over q}
\eqno(3.75)
$$
in physical units.
This simple result holds for any value of the transverse wavevector $q$.
Moreover, since it is a bulk property of the problem,
it also holds in the presence of an index mismatch,
just as the extinction lengths of the non-diffusive sectors for $Q=0$,
determined in section 3.1.4.

By inserting the explicit forms (3.72) into the expressions (3.74),
we obtain a $9\times 9$ linear system for the $Q$-dependent constants
$\{a_1, a_2, b_1, b_2, b_3, d_1, d_2, g_1, g_2\}$,
whose coefficients have complicated expressions
involving the functions $H(Q,s)$.
This system has been solved formally by means of the MACSYMA software:
the outcome for each constant contains thousands of products of up to seven
$H$-functions, so that this approach is of no practical use.
The above system is however easily solved numerically,
for any given value of $Q$.
This is the way we have chosen to follow for practical purposes.
\smallskip
\noindent{\bf 3.2.5. Summary of results}

We have achieved the exact analytical determination
of the enhanced backscattering cone in the absence of internal reflections.
Its dependence on polarisations is contained in five functions
of the reduced wavevector $Q$.
Two of them, $\g_{33}(Q)$ and $\g_{44}(Q)$, are given explicitly
in eqs. (3.62), (3.66),
while the other three, $\g_{11}(Q)$, $\g_{12}(Q)$, and $\g_{22}(Q)$,
are determined analytically,
up to a last step which consists in solving numerically a well-posed
$9\times 9$ linear system, for any fixed value of $Q$.

The following regimes are of special interest.

\noindent $\bullet$
For small $Q$, i.e., in the vicinity of the top of the cone,
$\g_{11}(Q)$, $\g_{12}(Q)$, and $\g_{22}(Q)$ have the common
characteristic triangular shape given by the general result (2.58),
while $\g_{33}(Q)$ and $\g_{44}(Q)$ have a smooth dependence on $Q^2$.
The enhancement factors and widths of the triangular cone,
for linear and circular polarisations,
have already been given in eqs. (3.41--44).

\noindent $\bullet$
For large $Q$, i.e., in the wings of the cone,
the leading contributions come from low-order scattering events, as usual.
The single-scattering contribution (2.45) is of little interest,
since it is subtracted in the formula (2.41) for the enhanced backscattering
peak.
The leading large-$Q$ behaviour of the enhancement factor
is thus given by double-scattering events.
The contribution of this class of events
can be obtained by solving eq. (3.47) to first order in the kernels $M_{pq}$.
As it turns out,
for large $Q$ these kernels become small, as expected,
but also local in the $\tau$-variable:
$M_{pq}(\tau,\tau')\approx m_{pq}(Q,0)\delta(\tau-\tau')$.
Furthermore, only the following kernels contribute to leading order in $1/Q$:
$$
\eqalign{
&m_{00}\approx{\pi\over 2Q},\quad m_{02}\approx{\pi\over 4Q},\quad
m_{04}\approx{3\pi\over 16Q},\quad m_{20}\approx{\pi\over 4Q},\quad
m_{22}\approx{\pi\over 16Q},\quad m_{40}\approx{3\pi\over 16Q},
}
\eqno(3.76)
$$
so that we are left with
$$
\eqalign{
\g_{11}(Q)\approx{3\over 4}\left(1+{3\pi\over 4Q}\right),\quad
\!\g_{22}(Q)\approx{3\over 4}\left(1+{9\pi\over 32Q}\right),\quad
\!\g_{33}(Q)\approx\g_{44}(Q)\approx-{3\over 4}\left(1+{3\pi\over 8Q}\right).
}
\eqno(3.77)
$$

We end up by illustrating a few interesting features of our results.
We first consider linearly polarised beams at normal incidence
and for parallel detection, in the following two geometries:

\noindent
(i) both polarisations parallel to $\Q$, i.e., $\psi_a=\psi_b=0$,

\noindent
(ii) both polarisations perpendicular to $\Q$, i.e., $\psi_a=\psi_b=\pi/2$.

\noindent
The enhancement factors, given by eq. (2.53), namely
$$
B_\l\ir{i}(Q)=1+{\g_{11}(Q)-3/4\over\g_{11}(0)},\quad
B_\l\ir{ii}(Q)=1+{\g_{22}(Q)-3/4\over\g_{11}(0)}
,\eqno(3.78)
$$
are plotted in Figure 3.
Both curves coincide with the result (3.41) at $Q=0$,
while they are slightly different from each other at $Q\ne 0$.
The enhancement factor of isotropic scattering of scalar waves,
after refs. [30, 19], is also shown for comparison.

Similarly, we consider linearly polarised beams
with perpendicular detection, in the following two cases:

\noindent
(iii) one polarisation parallel to $\Q$, i.e., $\psi_a=0$, $\psi_b=\pi/2$,

\noindent
(iv) both polarisations at $45^o$ with respect to $\Q$, i.e.,
$\psi_a=-\psi_b=\pi/4$.

\noindent
The enhancement factors,
$$
B_\r\ir{iii}(Q)=1+{\g_{44}(Q)-\g_{33}(Q)\over 2\g_{12}(0)},\quad
B_\r\ir{iv}(Q)=1+{\g_{11}(Q)+\g_{22}(Q)-2\g_{12}(Q)+2\g_{44}(Q)\over
4\g_{12}(0)}
,\eqno(3.79)
$$
are plotted in Figure 4.
Both factors coincide with the result (3.41) at $Q=0$,
while they are slightly different from each other at $Q\ne 0$.

We end up by considering circularly polarised beams at normal incidence.
The enhancement factors $B_1(Q)$ of the helicity-preserving channel,
and $B_{-1}(Q)$ of the channel of opposite helicity,
given by inserting the above results into
the general expressions (2.54), (2.63),
are plotted in Figures 5 and 6, respectively.
\smallskip
\noindent{\bf 4. LARGE INDEX MISMATCH REGIME}

Previous works [19--21] on multiple scattering of scalar waves
suggest that the SM equations cannot be solved analytically
in the presence of internal reflections,
which take place whenever the index mismatch $m=n/n_1$ is different from unity.
However, and interestingly enough, it has been shown in refs. [19--21]
that the problem becomes again tractable by analytical means
in the regime of a large index mismatch $(m\ll 1$ or $m\gg 1)$,
at least in the case of scalar waves.
The intuitive origin of this simplification is as follows.
If the ratio $m$ of both optical indices is very small
(respectively, very large), there is total reflection for almost any incidence
angle outside the medium (respectively, inside the medium),
except for a narrow cone around the normal incidence.
As a consequence, the radiation undergoes many internal reflections
at the boundary of the sample, and hence many scattering events,
before it can exit the medium.
In this section we extend this approach to the Rayleigh scattering
of electromagnetic waves.
\smallskip
\noindent{\bf 4.1. Diffuse reflection and transmission}

In this section we investigate the diffuse intensity in
reflection and in transmission in the large index mismatch regime.
To do so, it is advantageous to consider the matrix Green's function
$\Gr(\tau,\mu,\varphi,\tau',\mu',\varphi')$, which obeys eq. (2.13).
We anticipate on physical grounds
that only the $\varphi$-independent sector $(k=0)$ is important.
We thus rewrite eq. (2.13) as
$$
\eqalign{
\Gr\i{0}(\tau,\mu,&\tau',\mu')=\P\i{0}(\mu,\mu')\delta(\tau-\tau')\cr
&+\int_0^\tau\d\tau''\int_0^1{\d\mu''\over2\mu''}
e^{-(\tau-\tau'')/\mu''}\P\i{0}(\mu,\mu'').\Gr\i{0}(\tau'',\mu'',\tau',\mu')\cr
&+\int_\tau^{+\infty}\d\tau''\int_0^1{\d\mu''\over2\mu''}
e^{-(\tau''-\tau)/\mu''}\P\i{0}(\mu,-\mu'').
\Gr\i{0}(\tau'',-\mu'',\tau',\mu')\cr
&+\int_0^{+\infty}\d\tau''\int_0^1{\d\mu''\over2\mu''}
e^{-(\tau+\tau'')/\mu''}\P\i{0}(\mu,\mu'').\Gr\i{0}(\tau'',\mu'',\tau',\mu')\cr
&-\int_0^{+\infty}\d\tau''\int_0^1{\d\mu''\over2\mu''}
e^{-(\tau+\tau'')/\mu''}\bigl(\1-\R(\mu'')\bigr).\P\i{0}(\mu,\mu'').
\Gr\i{0}(\tau'',\mu'',\tau',\mu').\cr
}
\eqno(4.1)
$$

The above observations suggest to treat the small matrix $\1-\R(\mu)$ as a
perturbation.
In the limit of an infinite index mismatch $(m=0$ or $m=+\infty)$,
this matrix vanishes identically.
The rest of eq. (4.1), without the last line, has a zero mode of the form
$$
\M\nat=\I\nat\otimes\I\nat=\pmatrix{1&1&0&0\cr1&1&0&0\cr0&0&0&0\cr0&0&0&0\cr}
,\eqno(4.2)
$$
independent of $\tau$ and $\mu$.
This is demonstrated by the identity
$$
\int_{-1}^1{\d\mu\over2}\M\nat.\P\i{0}(\mu,\mu')=\M\nat
\quad\hbox{for all }\mu'
.\eqno(4.3)
$$

Along the lines of refs. [19--21], we expect that the Green's function
becomes proportional to the constant matrix $\M\nat$,
with a diverging prefactor, in the large index mismatch regime.
We thus look for a singular expansion of the form
$$
\Gr\i{0}(\tau,\mu,\tau',\mu')=C\M\nat+\Gr_0\i{0}(\tau,\mu,\tau',\mu')+\cdots
,\eqno(4.4)
$$
where it is understood that the constant $C$ diverges
as $m\to0$ or $m\to+\infty$, while $\Gr\i{0}_0$ stays finite,
and the dots stand for higher-order corrections.
We insert this expansion into eq. (4.1), and then act on both sides with
the operator $\int_0^{+\infty}\d\tau\int_{-1}^1(\d\mu/2)\M\nat$.
Integrals over the finite part $\Gr\i{0}_0$ of the Green's function
cancel out,
so that we are left with a simple expression for the constant $C$, namely
$$
C={2\over\ca{T}}
,\eqno(4.5)
$$
with
$$
\ca{T}={\ca{T}_\l+\ca{T}_\r\over 2},\quad
\ca{T}_\l=\int_0^1 2\mu\d\mu\,T_\l(\mu),\quad
\ca{T}_\r=\int_0^1 2\mu\d\mu\,T_\r(\mu)
.\eqno(4.6)
$$
The above quantities only depend on the index mismatch $m$.
They are interpreted as the mean flux transmission coefficients
of one boundary of the sample, averaged over incidence angles.
$\ca{T}_\l$ and $\ca{T}_\r$ correspond to prescribed polarisations,
while $\ca{T}$ is also averaged over both polarisation channels.

More explicitly, the expressions (2.8), (2.9)
of the Fresnel intensity coefficients allow us to perform
the integrals (4.6) in closed form, for both $m\ge 1$ and $m\le 1$.
It turns out that both cases can be gathered in the following formulas,
valid for $m\ge 1$:
$$
\eqalign{
m\ca{T}_\r(m)={1\over m}\ca{T}_\r\left({1\over m}\right)
&={4(2m+1)\over 3m(m+1)^2},\cr
m\ca{T}_\l(m)={1\over m}\ca{T}_\l\left({1\over m}\right)
&={2m(m^2-1)^2\over(m^2+1)^3}\ln{m+1\over m-1}\cr
&-{16m^3(m^4+1)\over(m^2-1)^2(m^2+1)^3}\ln m
+{4m^2(m^2+2m-1)\over(m^2-1)(m^2+1)^2}.\cr
}
\eqno(4.7)
$$
As expected, the flux transmission coefficients vanish in the regime
of a large index mismatch, according to
$$
\matrix{
m\to 0:\hfill&
\ca{T}_\l\approx 8m,\hfill&
\ca{T}_\r\approx\frd{8m}{3},\hfill&
\ca{T}\approx\frd{16m}{3},\cr
m\to+\infty:\hfill&
\ca{T}_\l\approx\frd{8}{m^3},\hfill&
\ca{T}_\r\approx\frd{8}{3m^3},\hfill&
\ca{T}\approx\frd{16}{3m^3}.\cr
}
\eqno(4.8)
$$

The knowledge of the matrix Green's function
yields by eqs. (2.16), (2.22--24) the following predictions for observables
of interest in the large index mismatch regime
$$
\eqalign{
\g\i{0}_{ij}(\mu_a,\mu_b)&\approx{2\mu_a\mu_b\over\ca{T}}\quad(i,j=1,2),\cr
\tau_{i}(\mu)&\approx{2\mu\over\ca{T}}\quad(i=1,2),\cr
\tau_0&\approx{4\over 3\ca{T}}.
}
\eqno(4.9)
$$
These leading-order results in the large index mismatch regime
are very similar to those obtained
in the case of isotropic [19, 20] and arbitrary anisotropic [21]
scattering of scalar waves.
Figure 7 shows plots of the mean transmission amplitudes $\ca{T}$, $\ca{T}_\l$,
and $\ca{T}_\r$, against the index mismatch $m$.
It is worth noticing that the reflection and transmission coefficients
for scalar waves coincide with $R_\r(\mu)$ and $T_\r(\mu)$,
so that $\ca{T}_\r$ was already involved in the
predictions of refs. [19--21] for scalar waves.

The behaviour of the quantities investigated above
in the large index mismatch regime involves,
besides the leading asymptotic behaviour in $1/\ca{T}$ derived above,
finite parts related to the Green's function $\Gr_0\i{0}$.
Moreover, $\Gr_0\i{0}$ also governs the (non-divergent)
large index mismatch behaviour of all the other quantities,
like e.g. the entries of the bistatic matrix $\g\i{0}_{ij}$
outside the (1,2) sector, or the bistatic matrices $\g\i{k}_{ij}$ for $k\ne 0$.
These finite parts cannot be determined analytically in general.
In the case of isotropic scattering of scalar waves,
the finite parts of some simple observables have been determined [19, 20],
either analytically or numerically.
\smallskip
\noindent{\bf 4.2. Enhanced backscattering cone}

We now extend the above analysis to the enhanced backscattering cone.
In analogy with the case of scalar waves, treated in refs. [19--21],
we want to show that the bistatic matrix $\g(Q)$ takes a simple scaling form
in the regime of small $Q$ and large index mismatch,
where enhanced backscattering is dominated by long-distance effects.

To do so, we look for a solution to the $Q$-dependent SM equations
(2.46) in the form
$$
\G\i{k}(\tau,\mu)\approx a(Q)e^{-Q\tau}\M\nat\delta_{k0}
,\eqno(4.10)
$$
for $m\ll 1$ or $m\gg 1$, and $Q\ll 1$.
The above ansatz is justified as follows:
the exponential fall-off in $\exp(-Q\tau)$ is quite general [see eq. (3.75)];
the $\varphi$-dependent sectors $(k\ne 0)$ are again neglected;
the proportionality to the matrix $\M\nat$
is assumed because of the structure (4.2), (4.3) of the zero mode
of the RTT problem at $Q=0$.

In analogy with section 4.1, we insert the form (4.10) into eq. (2.46),
and then act on both sides with the operator
$\int_0^{+\infty}\d\tau\int_{-1}^1(\d\mu/2)\M\nat$.
The integrals which do not involve the small matrix $\1-\R(\mu)$
can be performed exactly; their $Q$-dependence is given by
$m_{00}(Q,0)=m_0(iQ)=\arctan(Q)/Q\approx 1-Q^2/3$, by eq. (3.7).
The $Q$-dependence of the integral involving $\1-\R(\mu)$ can be neglected.
Consistently neglecting all corrections of order $Q^2$, we obtain
$$
\g_{ij}(Q)\approx{1\over\displaystyle{{2Q^{\vphantom{1}}\over 3}+{\ca{T}\over
2}}}\quad(i,j=1,2)
.\eqno(4.11)
$$
This prediction of the leading behaviour of the enhanced backscattering cone
in the large index mismatch regime is again very similar
to the case of scalar waves [19--21].

We end up by giving a few consequences of the above predictions
in the large index mismatch regime.
In the case of linear polarisations,
the enhancement factor (2.55) at the top of the cone reads $B(0)=2\cos^2\Psi$.
It assumes the maximal value $B_\Vert=2$ for parallel detection,
because the single-scattering contribution is negligible.
The width of the top of the cone (2.61) reads
$$
\Delta Q_\Vert\approx{3\ca{T}\over 4}
.\eqno(4.12)
$$
This last result also gives the width $\Delta Q_1$ of the top of the cone of
enhanced backscattering for the helicity-preserving channel
(2.66) in the case of circular polarisations.
\vfill\eject

\noindent{\bf 5. DISCUSSION}

In this paper we have extended to the Rayleigh scattering
of electromagnetic waves our previous investigations [19--21]
of multiple scattering of scalar waves in a thick-slab geometry.
Both the setup and the formalism of the present work
closely follow those references,
so that only the main lines of the derivations have been reproduced here.
The main advantage of this approach, based on RTT,
is that the role of skin layers,
and especially the effects of internal reflections,
are incorporated in a natural way.
The present approach has no phenomenological or approximate character,
besides the restriction of its validity to the regime $\lambda\ll\ell\ll L$,
in contrast with the widely used diffusion approximation.
Only few analytical results [17, 18] had been obtained
for electromagnetic waves along this line of thought,
since the pioneering work of Chandrasekhar [1].
It is, however, worth mentioning that the vector RTT formalism,
including the effects of internal reflections,
has been exposed earlier [31, 32], although these authors only solved
the SM equations numerically in some specific situations,
rather than investigating their general properties.
We have first derived general results on vector RTT
in section 2, where the mean values of observables are expressed
in terms of solutions to vector SM equations,
including the effects of polarisations and of internal reflections.
Closed-form expressions for these general predictions are then derived
in two cases, namely in the absence of internal reflections (in section
3), and in the regime of a large index mismatch (in section 4).

In the absence of internal reflections $(m=n/n_1=1)$,
the SM equations have been solved by means of the Wiener-Hopf technique.
We have presented in section 3 a self-contained exposition
of all the exact results known so far [1].
More importantly, we have given the first complete analytical derivation
of the cone of enhanced backscattering,
completing thus the analytical results of refs. [17, 18],
as well as some less accurate estimates,
obtained either by means of the diffusion approximation
or by numerical simulations [10, 11, 13].
As a general rule, illustrated by the first three items of Table 2,
quantities which are either averaged over the polarisation degrees
of freedom, or do not depend on polarisations at all,
are found to be very close to the corresponding figures
in the case of multiple isotropic scattering of scalar waves.
A similar observation has been made in ref. [21],
where various observables were compared for isotropic
and very anisotropic (forward) scattering of scalar waves.
The prototype of such quantities is the thickness $\tau_0$ of a skin layer,
expressed in units of the transport mean free path $\ell^*$.
This number is hardly sensitive to the anisotropy of the scattering mechanism
nor to polarisations: it always comes out to read $\tau_0\approx 0.71$ [4].

Other observables, such as the detailed shape of the cone
of enhanced backscattering, have a more or less pronounced dependence
on the polarisation channels of the incident and detected beams.
The last two items of Table 2 illustrate this point.
The first quantity under consideration
is the maximal enhancement factor, right at the top of the cone.
The value $B_\l$ of eq. (3.41),
corresponding to linear polarisations and parallel detection,
as well as the value $B_1$ of eq. (3.43),
corresponding to circular polarisations and detection in the channel
of same helicity, are compared to the analogous result for scalar waves
with isotropic scattering [30, 19], denoted by $B$:
the figures are definitely different from each other,
although relatives differences are less than 10\%.
Second, the width of the triangular top of the
backscattering cone is considered.
The value $\Delta Q_\l$ of eq. (3.42),
corresponding to linear polarisations and parallel detection,
as well as the value $\Delta Q_1$ of eq. (3.44),
corresponding to circular polarisations and detection in the channel
of same helicity, are compared to the analogous result for scalar waves
with isotropic scattering [30, 19], denoted by $\Delta Q$:
relative differences are more important in this case, going up to some 40\%.
Finally, so far there are essentially no analytical results concerning
the RTT approach
to the general problem of multiple scattering of electromagnetic waves,
taking into account the combined effects of anisotropic scattering
and polarisations.
We can infer from the results of ref. [21]
on the multiple scattering of scalar waves
that the anisotropy of the scattering mechanism will have
little residual effects,
once the principal scaling is taken into account by expressing
observables in terms of the transport mean free path $\ell^*$.

In the presence of internal reflections $(m=n/n_1\ne 1)$,
analytical predictions for the various observables of interest
have been derived in the large index mismatch regime $(m\ll 1$ or $m\gg 1)$,
along the lines of previous investigations of scalar waves,
with isotropic [19, 20] and arbitrary anisotropic [21] scattering.
The effects of internal reflections have been studied [22--25]
using several variants of the diffusion approximation.
Ref. [26] provides a recent overview of these approaches to the subject.
Within the framework of RTT,
the drastic simplification which takes place in the large index mismatch regime
has a clear intuitive explanation.
Since the transmission through the boundaries of the sample is small,
radiation is reinjected many times before it can leave the medium.
As a consequence, the skin layers become very thick and,
more importantly, the radiation field is uniform over most of these layers.
The results (4.9) turn out to have the very same form
as for multiple scattering of scalar waves,
either with isotropic [19, 20] or very anisotropic [21] scattering.
Most certainly, the very same analytical forms also hold true
in the more general case of multiple scattering of electromagnetic waves,
including both anisotropy and polarisation effects,
and they are expected to provide an overall satisfactory description
of the full dependence of physical quantities on the index mismatch,
especially in the range of most interest $(m\ge 1)$.

The present investigations of multiple
Rayleigh scattering of electromagnetic waves,
as well as the previous ones on isotropic and anisotropic
scattering of scalar waves [19--21],
have so far only dealt with the mean intensity,
averaged over the random positions of the scatterers in the sample.
For any given sample of scattering medium, however,
the intensity has strong fluctuations,
which manifest themselves as speckles.
For instance, the probability law of the fluctuating intensity
at a given point is known as Rayleigh's law:
$p(I)\sim\exp(-I/\langle I\rangle)$.
The generalisation of Rayleigh's law to polarised radiation is
fully characterised by the four Stokes parameters [33].
Various correlation functions, aiming at a more detailed description
of intensity fluctuations and speckle patterns,
have been the subject of recent theoretical
and experimental investigations [34].
We mention the extension of these results,
in order to include polarisation effects, as an interesting open problem.
\medskip
\noindent{\bf Acknowledgements}

The work of Th.M.N. has been supported by the
Royal Dutch Academy of Arts and Sciences (KNAW).

\vfill\eject
{\parindent 0pt
{\bf REFERENCES}
\bigskip

[1] S. Chandrasekhar, {\it Radiative Transfer} (Dover, New-York, 1960).

[2] V.V. Sobolev, {\it A Treatise on Radiative Transfer} (Van Nostrand,
Princeton, N.J., 1963).

[3] A. Ishimaru, {\it Wave Propagation and Scattering in Random Media}, in 2
volumes (Academic, New-York, 1978).

[4] H.C. van de Hulst, {\it Multiple Light Scattering}, in 2 volumes (Academic,
New-York, 1980).

[5] Y. Kuga and A. Ishimaru, J. Opt. Soc. Am. A {\bf 1}, 831 (1984); M.P. van
Albada and A. Lagendijk, Phys. Rev. Lett. {\bf 55}, 2692 (1985); P.E. Wolf and
G. Maret, Phys. Rev. Lett. {\bf 55}, 2696 (1985).

[6] M.B. van der Mark, M.P. van Albada, and A. Lagendijk, Phys. Rev. B {\bf
37}, 3575 (1988).

[7] A. Ishimaru and L. Tsang, J. Opt. Soc. Am. A {\bf 5}, 228 (1988).

[8] G.C. Papanicolaou and R. Burridge, J. Math. Phys. {\bf 16}, 2074 (1975).

[9] K.M. Watson, J. Math. Phys. {\bf 10}, 688 (1969).

[10] M.J. Stephen and G. Cwilich, Phys. Rev. B {\bf 34}, 7564 (1986); G.
Cwilich and M.J. Stephen, Phys. Rev. B {\bf 35}, 6517 (1987).

[11] E. Akkermans, P.E. Wolf, R. Maynard, and G. Maret, J. Phys. (France) {\bf
49}, 77 (1988).

[12] F.C. MacKintosh and S. John, Phys. Rev. B {\bf 37}, 1884 (1988); Phys.
Rev. B {\bf 40}, 2383 (1989).

[13] M.P. van Albada and A. Lagendijk, Phys. Rev. B {\bf 36}, 2353 (1987).

[14] B.A. van Tiggelen, R. Maynard, and Th.M. Nieuwenhuizen, Phys. Rev. B {\bf
53}, 2881 (1996).

[15] B.A. van Tiggelen, Phys. Rev. Lett. {\bf 75}, 422 (1995).

[16] K.J. Peters, Phys. Rev. B {\bf 46}, 801 (1992).

[17] M.I. Mishchenko, Phys. Rev. B {\bf 44}, 12597 (1991); J. Opt. Soc. Am. A
{\bf 9}, 978 (1992).

[18] V.D. Ozrin, Waves in Random Media {\bf 2}, 141 (1992).

[19] Th.M. Nieuwenhuizen and J.M. Luck, Phys. Rev. E {\bf 48}, 569 (1993).

[20] Th.M. Nieuwenhuizen, {\it Veelvoudige verstrooing van golven}, unpublished
lecture notes in Dutch (University of Amsterdam, 1993).

[21] E. Amic, J.M. Luck, and Th.M. Nieuwenhuizen, J. Phys. A {\bf 29},
4915 (1996).

[22] A. Lagendijk, R. Vreeker, and P. de Vries, Phys. Lett. A {\bf 136}, 81
(1989).

[23] I. Freund and R. Berkovits, Phys. Rev. B {\bf 41}, 496 (1990).

[24] J.X. Zhu, D.J. Pine, and D.A. Weitz, Phys. Rev. A {\bf 44}, 3948 (1991).

[25] I. Freund, Phys. Rev. A {\bf 45}, 8854 (1992).

[26] I. Freund, J. Opt. Soc. Am. A {\bf 11}, 3274 (1994).

[27] M. Born and E. Wolf, {\it Principles of Optics} (Pergamon, 1965).

[28] D.S. Wiersma, M.P. van Albada, B.A. van Tiggelen, and A. Lagendijk, Phys.
Rev. Lett. {\bf 74}, 4193 (1995).

[29] B.A. van Tiggelen, D.S. Wiersma, and A. Lagendijk, Europhys. Lett. {\bf
30}, 1 (1995).

[30] E.E. Gorodnichev, S.L. Dudarev, and D.B. Rogozkin, Phys. Lett. A {\bf
144}, 48 (1990).

[31] L. Tsang and J.A. Kong, Radio Sci. {\bf 13}, 763 (1978); R.T. Shin and
J.A. Kong, J. Appl. Phys. {\bf 52}, 4221 (1981).

[32] Q. Ma and A. Ishimaru, I.E.E.E. Trans. Antennas and Propagation {\bf 39},
1626 (1991); C.M. Lam and A. Ishimaru, I.E.E.E. Trans. Antennas and Propagation
{\bf 41}, 851 (1993).

[33] L. Mandel, Proc. Phys. Soc. {\bf 81}, 1104 (1963), and references therein.

[34] M.C.W. van Rossum, J.F. de Boer, and Th.M. Nieuwenhuizen, Phys. Rev. E
{\bf 52}, 2053 (1995), and references therein.

\vfill\eject
{\bf CAPTIONS FOR TABLES AND FIGURES}
\bigskip
{\bf Table 1:}
Definitions and notations for kinematic and other useful quantities.

{\bf Table 2:}
Comparison of various quantities of interest, defined in section 2,
from the known exact solutions in the absence of internal reflections.
First row: isotropic scattering of scalar waves, after ref. [19].
Second row: Rayleigh scattering of electromagnetic waves
(section 3 of this work).
Third row: relative difference of second case with respect to first one.

{\bf Table 3:}
Dimensionless reduced extinction lengths of the various polarised components
of the diffuse intensity.
First row: exact extinction lengths, deduced in section 3.1.4
from the solution to the SM equation.
Second row: approximate extinction lengths, obtained by means of the diffusion
approximation [10].
Third row: relative difference of second case with respect to first one.
Fourth row: notations used in ref. [10].

\bigskip
{\bf Figure 1:} Plot of angular dependence of transmitted intensity
in the absence of internal reflections, against $\mu=\cos\theta$.
Full lines: $\tau_1(\mu)$ (lower curve) and $\tau_2(\mu)$ (upper curve),
corresponding to both polarisations channels for Rayleigh scattering
(this work).
Dashed line: $\tau\sca(\mu)/2$, corresponding to
half the result for isotropic scattering of scalar waves, after ref. [19].

{\bf Figure 2:} Plot of angular dependence of degree of polarisation
$P(\mu)$ of diffuse transmitted intensity,
in the absence of internal reflections,
against $\mu=\cos\theta$.

{\bf Figure 3:} Plot of enhancement factors $B_\l(Q)$
for linearly polarised beams at normal incidence and parallel detection,
in the absence of internal reflections, in two cases defined in the text.
Upper full line: case (i) ($\psi_a=\psi_b=0$).
Lower full line: case (ii) ($\psi_a=\psi_b=\pi/2$).
Dashed line: same quantity for isotropic scattering
of scalar waves, after ref. [19].

{\bf Figure 4:} Plot of enhancement factors $B_\r(Q)$
for linearly polarised beams at normal incidence and perpendicular detection,
in the absence of internal reflections, in two cases defined in the text.
Lower full line: case (iii) ($\psi_a=0$, $\psi_b=\pi/2$).
Upper full line: case (iv) ($\psi_a=-\psi_b=\pi/4$).

{\bf Figure 5:} Plot (full line) of enhancement factor $B_1(Q)$
for circularly polarised beams at normal incidence
in the helicity-preserving channel, in the absence of internal reflections.
Dashed line: same quantity for isotropic scattering
of scalar waves, after ref. [19].

{\bf Figure 6:} Plot of enhancement factor $B_{-1}(Q)$
for circularly polarised beams at normal incidence
in the channel of opposite helicity, in the absence of internal reflections.

{\bf Figure 7:} Plot of mean flux transmission coefficients,
against optical index mismatch $m$.
Upper dashed line: $\ca{T}_\l(m)$.
Lower dashed line: $\ca{T}_\r(m)$ (also corresponds to scalar waves).
Full line: their average $\ca{T}(m)$.

\vfill\eject
}
\centerline {\bf Table 1}
\smallskip
$$\vbox{\init\halign to 16truecm
{\strut#&\vrule#\tabskip=1em plus 2em&
\hfil$#$\hfil&\vrule#&\hfil$#$\hfil&\vrule#&\hfil$#$\hfil&
\vrule#\tabskip 0pt\crr
&&\ &&\ &&\ &\cr
&&\ &&\hbox{outside medium} &&\hbox{inside medium} &\cr
&&\ &&\ &&\ &\crr
&&\ &&\ &&\ &\cr
&&\hbox{optical index} &&n_1 &&n=m n_1 &\cr
&&\ &&\ &&\ &\crr
&&\ &&\ &&\ &\cr
&&\hbox{wavenumber} &&k_1=n_1\omega/c=2\pi/\lambda_1
&&k=n\omega/c=2\pi/\lambda &\cr
&&\ &&\ &&\ &\crr
&&\ &&\ &&\ &\cr
&&\hbox{incidence angle} &&\theta_1 &&\theta&\cr
&&\ &&\ &&\ &\cr
&&\ &&
\matrix{\cos\theta_1=\sqrt{1-m^2\nu^2}\hfill\cr\sin\theta_1=m\nu\hfill\cr}
&&\matrix{\cos\theta=\mu\hfill\cr\sin\theta=\nu=\sqrt{1-\mu^2}\hfill\cr}&\cr
&&\ &&\ &&\ &\crr
&&\ &&\ &&\ &\cr
&&\hbox{parallel wavevector} && p=k_1\cos\theta_1 && P=k\cos\theta &\cr
&&\ &&\ &&\ &\crr
&&\ &&\ &&\ &\cr
&&
\matrix{\hbox{total reflection}\cr\hbox{condition}} &&
\matrix{m<1\ \ \hbox{and}\ \ \sin\theta_1>m\cr (\hbox{i.e.}\ \ P\ \
\hbox{imaginary})} &&
\matrix{m>1\ \ \hbox{and}\ \ \sin\theta>1/m \cr (\hbox{i.e.}\ \ p\ \
\hbox{imaginary})} &\cr
&&\ &&\ &&\ &\crr
}}$$
$$\vbox{\init\halign to 16truecm
{\strut#&\vrule#\tabskip=1em plus 2em&
\hfil$#$\hfil&\vrule#&\hfil$#$\hfil&
\vrule#\tabskip 0pt\crr
&&\ &&\ &\cr
&&\hbox{transverse wavevector}&&
\vert{\bf q}\vert=q=k_1\sin\theta_1=k\sin\theta=k\nu &\cr
&&\ &&\ &\crr
&&\ &&\ &\cr
&&\hbox{azimuthal angle} &&\varphi &\cr
&&\ &&\ &\crr
}}$$
\vfill\eject

\centerline {\bf Table 2}
\smallskip
$$\vbox{\init\halign to 16truecm
{\strut#&\vrule#\tabskip=1em plus 2em&
\hfil$#$\hfil&\vrule#&
\hfil$#$\hfil&\vrule#&
\hfil$#$\hfil&\vrule#&
\hfil$#$\hfil&\vrule#\tabskip 0pt\crr
&&\ &&\ &&\ &&\ &\cr
&&\ &&\dbl{isotropic scattering}{(scalar waves)}&&\dbl{Rayleigh scattering}
{(electromagnetic waves)}&&\Delta(\%)&\cr
&&\ &&\ &&\ &&\ &\crr
&&\ &&\ &&\ &&\ &\cr
&&\dbl{skin layer}{thickness}&&\tau_0=0.710\,446&&\tau_0=0.712\,110&&0.23&\cr
&&\ &&\ &&\ &&\ &\crr
&&\ &&\ &&\ &&\ &\cr
&&\dbl{diffuse}{transmission}&&{1\over 2}\tau\sca(1)=2.518\,237
&&\tau_i(1)=2.538\,761\;\;{\scriptstyle(i=1,2)}&&0.81&\cr
&&\ &&\ &&\ &&\ &\crr
&&\ &&\ &&\ &&\ &\cr
&&\dbl{diffuse}{reflection}&&\g(1,1)=4.227\,681&&
\g_{11}+\g_{12}=4.588\,369&&8.5&\cr
&&\ &&\ &&\ &&\ &\crr
&&\ &&\ &&\ &&\ &\cr
&&\dbl{enhancement}{at top of cone}&&B=1.881\,732&&
\matrix{B_\l=1.752\,088\cr\cr B_1=2\cr}
&&\matrix{-6.9\cr\cr 6.3\cr}&\cr
&&\ &&\ &&\ &&\ &\crr
&&\ &&\ &&\ &&\ &\cr
&&\dbl{width}{of top of cone}&&\Delta Q=1/2&&
\matrix{\Delta Q_\l=0.704\,063\cr\cr\Delta Q_1=0.407\,487\cr}
&&\matrix{41\cr\cr -19\cr}&\cr
&&\ &&\ &&\ &&\ &\crr
}}$$
\vfill\eject

\centerline {\bf Table 3}
\smallskip
$$\vbox{\init\halign to 16truecm
{\strut#&\vrule#\tabskip=1em plus 2em&
\hfil$#$\hfil&\vrule#&
\hfil$#$\hfil&\vrule#&
\hfil$#$\hfil&\vrule#&
\hfil$#$\hfil&\vrule#\tabskip 0pt\crr
&&\ &&\ &&\ &&\ &\cr
&&\hbox{exact}&&\hbox{diffusion approximation}&&\Delta(\%)&&
\dbl{notation}{of ref. [10]}&\cr
&&\ &&\ &&\ &&\ &\crr
&&\ &&\ &&\ &&\ &\cr
&&\ell_1=1&&\ell_1\dif=\sqrt{\frd{1}{5}}=0.447\,213&&-55&&\ell'_3&\cr
&&\ &&\ &&\ &&\ &\crr
&&\ &&\ &&\ &&\ &\cr
&&\ell_3=1.093\,116&&\ell_3\dif=\sqrt{\frd{13}{21}}=0.786\,796&&
-28&&\ell'_2&\cr
&&\ &&\ &&\ &&\ &\crr
&&\ &&\ &&\ &&\ &\cr
&&\ell_4=1.349\,587&&\ell_4\dif=\sqrt{\frd{23}{21}}=1.046\,536&&-22&&\ell_2&\cr
&&\ &&\ &&\ &&\ &\crr
&&\ &&\ &&\ &&\ &\cr
&&\ell_5=1.172\,669&&\ell_5\dif=\sqrt{\frd{3}{5}}=0.774\,596&&-34&&\ell_3&\cr
&&\ &&\ &&\ &&\ &\crr
}}$$
\bye